\documentclass[11pt,a4paper]{article}

\usepackage{jheppub}
\usepackage[utf8]{inputenc}
\usepackage{amsmath,xcolor}
\usepackage{multirow, graphicx,amssymb,url,mathrsfs,amsmath}
\usepackage{wrapfig,boxedminipage,setspace,subcaption,epsfig}
\usepackage{amsxtra,amstext,latexsym,dsfont,amsfonts}
\usepackage{color,eucal}
\usepackage{float}
\usepackage{slashed,comment}
\usepackage{kotex}
\usepackage{tensor}
\usepackage{indentfirst}

\title{Holographic geometry/real-space entanglement correspondence and metric reconstruction\\} 

\author[a,b]{Xuanting Ji,} 
\author[b]{Xin-Xiang Ju,}
\author[b,c]{Ya-Wen Sun,}
\author[d,e]{Yuan-Tai Wang,}
\author[b]{He-Lin Zhou}

\emailAdd{jixuanting@cau.edu.cn}
\emailAdd{juxinxiang21@mails.ucas.ac.cn}
\emailAdd{yawen.sun@ucas.ac.cn}
\emailAdd{wangyuantai@ustc.edu.cn}
\emailAdd{zhouhelin20@mails.ucas.ac.cn}

\affiliation[a]{Department of Applied Physics, College of Science, China Agricultural University, Beijing 100083, China}
\affiliation[b]{School of Physical Sciences, University of Chinese Academy of Sciences, Beijing 100049, China}
\affiliation[c]{Kavli Institute for Theoretical Sciences, University of Chinese Academy of Sciences, Beijing 100049, China}
\affiliation[d]{Interdisciplinary Center for Theoretical Study, University of Science and Technology of China, Hefei, Anhui 230026, China}
\affiliation[e]{Peng Huanwu Center for Fundamental Theory, Hefei, Anhui 230026, China  
}

\abstract{
In holography, the boundary entanglement structure is believed to be encoded in the bulk geometry. In this work, we investigate the precise correspondence between the boundary real-space entanglement and the bulk geometry. By the boundary real-space entanglement, we refer to the conditional mutual information (CMI) for two infinitesimal subsystems separated by a distance $l$, and the corresponding bulk geometry is at a radial position $z_*$, namely the turning point of the entanglement wedge for a boundary region with a length scale $l$. In a generic geometry described by a given coordinate system, $z_*$ can be determined locally by $l$, while the exact expression for $z_*(l)$ depends on the gauge choice, reflecting the inherent nonlocality of this seemingly local correspondence. We propose to specify the function $z_*(l)$ as the criterion for a gauge choice, and with the specified gauge function, we verify the exact correspondence between the boundary real-space entanglement and the bulk geometry. Inspired by this correspondence, we propose a new method of bulk metric reconstruction from boundary entanglement data, namely the CMI reconstruction. In this CMI proposal, with the gauge fixed a priori by specifying $z_*(l)$, the bulk metric can be reconstructed from the relation between the bulk geometry and the boundary CMI. The CMI reconstruction method establishes a connection between the differential entropy prescription and Bilson's general algorithm for metric reconstruction.}

\begin{document}


\maketitle

\section{Introduction}\label{1}

The AdS/CFT correspondence \cite{Maldacena:1997re} has established a deep connection between the geometry of a gravitational spacetime and the quantum information properties of the boundary quantum system. Specifically, the geometry of the bulk is intricately linked to the boundary entanglement structure. The entanglement entropy of a boundary subregion is found to be proportional to the area of the corresponding minimal surface in the bulk. This relationship, known as the Ryu-Takayanagi formula \cite{Ryu:2006bv}, provides a concrete example of how geometric notions in the bulk can be translated into quantum information quantities in the boundary. Moreover, the concept of subregion subregion duality \cite{Czech:2012bh,Wall:2012uf,Headrick:2014cta,Bousso:2022hlz,Espindola:2018ozt,Saraswat:2020zzf,dong2016reconstruction} and the subregion subalgebra duality later developed in \cite{Leutheusser:2022bgi} further connects a bulk subregion with a boundary subsystem. 

In \cite{Ju:2024xcn}, it was proposed that the bulk geometry at different radial scales corresponds to the boundary entanglement structure at corresponding real-space scales in AdS$_3$/CFT$_2$ {inspired by observations in \cite{Ju:2023bjl,Ju:2023dzo,Balasubramanian:2013rqa}}. To be more precise, the boundary entanglement structure at the real-space scale $l$ is characterized by the conditional mutual information (CMI) between two infinitesimal boundary subsystems of the same lengths at a distance $l$, with the interval between these two subsystems serving as the condition. It is proposed to be determined by the bulk geometry at a radial scale $z_*$ determined by $l$ \cite{Ju:2024xcn}. 

CMI is a physical quantity that quantifies the correlations between two quantum systems conditioned on a third system. Recently, it has been found that CMI characterizes the entanglement of subsystems separated at a distance by incorporating the multipartite entanglement \cite{Ju:2024xcn,Ju:2024hba,Ju:2024kuc} that the two subsystems participate with a third subsystem. It provides more physical information on the entanglement structures of two subsystems compared to the mutual information. 
{Another merit of CMI in evaluating the entanglement properties is that it is independent of the UV cutoff. This is a significant property superior to the entanglement entropy which is cutoff-dependent and diverges as the cutoff goes to zero.}
This bulk geometry/boundary CMI correspondence was confirmed in pure AdS and in certain geometries with modified IR geometry, where in the latter case the long range entanglement structure is altered due to the change in the IR geometry \cite{Ju:2024xcn}. Such kind of correspondence was also detected in the bulk metric reconstruction in \cite{Xu:2023eof}.

{This bulk geometry/boundary real space entanglement correspondence could be viewed as a real space manifestation of the UV/IR relation in AdS/CFT. The usual UV/IR relation in the dictionary \cite{Peet:1998wn,Papadimitriou:2004ap} encapsulates how energy scales in the boundary CFT are geometrically encoded in the radial direction of the bulk spacetime. Specifically, UV physics in the boundary theory corresponding to short-distance correlations is associated with the near-boundary region of the bulk, while IR physics linked to long-distance behavior is encoded in the deep interior of AdS. Here in our work, we emphasize the real space scale version of the UV/IR relation, i.e. the bulk geometry at a certain radial scale determines the boundary quantum entanglement behavior at a corresponding real space distance scale. Due to the uncertainty relation, this real space version of the UV/IR relation cannot be a local one, therefore it only holds in a given gauge.}

However, due to the diffeomorphism invariance of gravity, the correspondence between the local geometry and the boundary entanglement structure should be gauge dependent, i.e. the exact {form} of the function $z_*(l)$ varies in different coordinate systems. This indicates that while this exact correspondence between the real-space entanglement structure and the bulk geometry at a certain radial position indeed holds, the explicit  corresponding bulk radial scale should be determined by the boundary real-space scale differently in different coordinate systems.

Generically, a physical rule should be coordinate-independent (gauge-independent), that is, it holds universally. Therefore, in a holographic theory, the correspondence between the emergent bulk geometry and the boundary entanglement is conjectured to exist naturally without imposing any gauge choice since the rules and phenomena behind these quantities are physical. However, generally in a gravitational theory, the diffeomorphism invariance results in gauge degrees of freedom, relativity of local spacetime structures and non-local causality. As a consequence, the seemingly local correspondence between the bulk radial position and the boundary scale, in our case, cannot be uncovered unless a specific gauge is chosen since these quantities are coordinate-dependent (gauge-dependent).

To understand this point more precisely, in this work we will directly calculate the dual CMI and show that its behavior at a distance $l$ is indeed determined by the geometry at a corresponding radial position $z_*(l)$, where, however, $z_*(l)$ has to be first calculated from an integration of the geometry from the boundary up to $z_*$. This means that though the boundary entanglement structure at length $l$ is determined by a bulk local geometry at $z_*$, we do not know in prior which $z_*$ should correspond to a given $l$ because {in order} to get the relation $z_*(l)$, we need to integrate over the geometry in the whole UV region up to $z=z_*$. Therefore, the behavior of CMI should still secretly depend on the geometry in the {exterior} UV region {rather than} only the geometry at $z=z_*$.  

To address this problem of not knowing the relation $z_*(l)$ in advance when we {attempt} to establish an exact correspondence between the bulk geometry at a radial scale and the boundary real-space entanglement structure, we draw inspiration from the argument that the relation $z_*(l)$ should be gauge-dependent. We adopt an alternative approach that allows us to solve the problem from the opposite direction: instead of starting with a given coordinate system and solving $z_*(l)$, we propose to use the function $z_*(l)$ to determine the gauge choice. By directly specifying a function of $z_*(l)$, we establish it as {an} a prior criterion for fixing the gauge choice of the coordinate system. That is, the gauge choice is determined from this $z_*(l)$ function.

As an example, we will start with the simplest gauge $z_*=l$ and work out the explicit dependence of the boundary CMI at a distance $l$ on the bulk geometry at the corresponding radial position $z_*=l$. For other coordinate choices, there could likewise be relations between the boundary CMI at a distance $l$ and bulk metric components at a radial position determined by a different function $z_*(l)$. This relation will depend on the specific choice of the coordinate system. {Furthermore,} this boundary entanglement structure/bulk radial geometry relation could in principle be generalized to the case of the CMI for two infinitesimal boundary subsystems with different lengths.

This finding stimulates us to propose a new method for the reconstruction of bulk metric from boundary entanglement data, or more precisely, CMI data. Holographic spacetime reconstruction leverages the AdS/CFT correspondence to derive bulk geometry from boundary information. As a main focus, the reconstruction goal has been achieved in typical geometric backgrounds using holographic entanglement entropy. The RT prescription identifies the entanglement entropy of a boundary subregion with the area of the bulk minimal surface homologous to that subregion, establishing the entanglement entropy as a geometric probe \cite{Ryu:2006bv}. Notably, Bilson developed the reconstruction algorithm for the spacetime metric in asymptotically AdS geometry \cite{Bilson:2008ab,Bilson:2010ff}. Another bulk reconstruction method is the differential entropy reconstruction, where bulk points and distances can be reconstructed from boundary entanglement data \cite{Hubeny:2014qwa,Headrick:2014eia,Myers:2014jia,Czech:2014wka,Balasubramanian:2018uus}. The algorithm was extended 
by employing deep learning which is a form of machine learning utilizing deep neural networks \cite{Park:2022fqy,Park:2023slm,Ahn:2024jkk}. Similar methods were used to derive the bulk metric from boundary optical conductivity \cite{Ahn:2024gjf}. The reconstruction was also extended to the spacetimes slightly breaking the translational symmetry using perturbative PDE methods \cite{Jokela:2025ime}. 
Alternatively, {the bulk geometry was demonstrated using the entanglement entropy in CFT} \cite{Wang:2018vbw}, and an algorithm of the reconstruction from fine-grained entanglement entropy was developed using holographic bit threads \cite{Agon:2020mvu}. {For a formal discussion on fixing the metric from boundary regions, one can refer to} \cite{Alexakis2020,Bao:2020abm}. See also \cite{Hammersley:2006cp,Hammersley:2007ab,Spillane:2013mca,Bao:2019bib,Jokela:2020auu,Cao:2020uvb,Jokela:2023rba,Xu:2023eof} for relevant proposals of the metric reconstruction.

In addition to the trials of entanglement entropy, methods including subregion duality and entanglement wedge reconstruction were developed, mapping boundary entanglement structures to bulk connectivity \cite{Jafferis:2015del,Dong:2016eik}. Quantum error correction frameworks and tensor network models were also applied, showing how bulk locality and geometry could emerge from boundary entanglement \cite{Almheiri:2014lwa,Pastawski:2015qua,Hayden:2016cfa}. Other successful reconstruction approaches also attract attention, such as modular Hamiltonians \cite{Roy:2018ehv,Kabat:2018smf}, boundary light-cone cuts \cite{Engelhardt:2016wgb,Engelhardt:2016crc,Hernandez-Cuenca:2020ppu}, correlation functions in excited states \cite{Caron-Huot:2022lff}, Wilson loops \cite{Hashimoto:2020mrx}, and holographic complexities \cite{Hashimoto:2021umd,Xu:2023eof}.

Our new method of the bulk metric reconstruction, named the CMI reconstruction proposal, is based on the $z_*(l)$ relation in a given gauge (coordinate choice). We specify the form of $z_*(l)$, and then obtain the corresponding bulk geometry at the radial position $z_*$ from boundary CMI at distance $l$ using the specified relation $z_*(l)$.
This step of the reconstruction is performed in the specific coordinate system specified by the gauge choice of $z_*(l)$. Note that this coordinate system can be viewed as a particular gauge choice in Bilson's reconstruction method \cite{Bilson:2008ab,Bilson:2010ff}. With the geometry obtained in this coordinate system, we can perform coordinate transformations to any other coordinate systems to obtain the geometry in those coordinate systems. 

{The advantage of the CMI metric reconstruction lies in several aspects. First, unlike the entanglement entropy in quantum field theory, CMI is inherently cutoff-independent, ensuring the cancellation of all UV divergences. This eliminates scheme-dependent ambiguities, providing a robust framework for the bulk metric reconstruction. Second, with the CMI prescription of metric reconstruction, the bulk radial coordinate appears to have a local mapping to the boundary length scale resulting from the specific gauge choice (the radial coordinate transformation), which manifests the relationship of bulk radial geometry and boundary physics. Therefore, the CMI metric reconstruction method establishes a more geometrically transparent connection between bulk and boundary physics, providing a clearer and more intuitive physical interpretation.  
Finally, this method also greatly simplifies the integration for the metric reconstruction. 
With high computational efficiency, the CMI prescription provides accurate reconstruction formulae and does not rely on a specific form of the bulk metric or any boundary conditions.}

This procedure can be extended to the scenario where the boundary CMI involves two infinitesimal subsystems of different lengths. In such a case, different bulk metric components are determined from the boundary CMI at a distance $l$, which can also be directly worked out. Then the differential entropy reconstruction of the geometry at a bulk point \cite{Czech:2014ppa,Czech:2014tva,Burda:2018rpb} can be viewed as an integration version of the CMI reconstruction {over} different ratios of the lengths of the two boundary infinitesimal subsystems. Each CMI reconstruction in the integration is gauge dependent while after this integration the reconstructed quantity is gauge invariant. Consequently, the differential entropy reconstruction is gauge invariant while the CMI reconstruction can be viewed as a decomposition into each gauge choice. Therefore, this CMI reconstruction method builds a connection between the methods of the differential entropy reconstruction \cite{Czech:2014ppa,Czech:2014tva,Burda:2018rpb} and the Bilson's reconstruction method \cite{Bilson:2008ab,Bilson:2010ff}, {providing insights into} the consistent relationship between these two methods.

The rest of the paper is organized as follows. In section \ref{2}, we will present the explicit calculation of the relationship between the boundary CMI at a length $l$ and the bulk geometry. We will also introduce a specific gauge choice and the corresponding CMI-radial scale correspondence, providing several explicit examples for illustration.
In section \ref{3}, we will first review Bilson's reconstruction method and the differential entropy reconstruction method. Then we will introduce our explicit algorithm for the CMI bulk metric reconstruction.
Section \ref{4} will provide several examples of the CMI reconstruction process. Section \ref{5} is devoted to conclusion and discussion.

\section{The radial geometry/boundary CMI relation}\label{2}

In holography, there is a deep connection between the bulk geometry and boundary entanglement structure. It has been found that the bulk IR geometry affects the boundary long range entanglement behavior \cite{Balasubramanian:2013rqa,Balasubramanian:2013lsa,Ju:2023bjl, Ju:2023dzo}, where the long range entanglement is characterized by the behavior of the CMI between two distant small subregions \cite{Ju:2024xcn}. Removing or modifying the IR geometry would result in a loss or a modification of long range entanglement. In this section, we hope to understand the exact correspondence between the bulk geometry at different radial positions and the boundary CMI with a direct calculation of the holographic entanglement entropy and the CMI. We will show that this exact correspondence indeed exists and that the relation between the radial scale and the boundary distance scale depends on gauge choices of coordinate systems. 

\subsection{Holographic entanglement entropy and conditional mutual information}

To start with, first we briefly review the widely used entanglement measures in quantum information theory. The entanglemen entropy is the most representative measure of bipartite entanglement in a pure state quantum system. For a subsystem $A$, the corresponding entanglement entropy $S(A)$ is defined by the von Neumann entropy
\begin{equation}\label{EE}
\begin{split}
\begin{aligned}
    S(A)&= -\mathrm{Tr}_A\; \rho_A\log\rho_A,\\
    \rho_A&=\mathrm{Tr}_{\Bar{A}}\; \rho,
\end{aligned}
\end{split}
\end{equation}
where $\rho_A$ is the reduced density matrix of $A$ from the density matrix $\rho$ of the whole system. Whereas the entanglement entropy is no longer a good entanglement measure for mixed-state systems {because it would mix classical correlations and quantum entanglement}. To overcome this problem, a well-studied mixed-state generalization of the entanglement entropy for bipartite systems ($A$ and $B$) is the mutual information (MI) $I(A,B)$
\begin{equation}\label{MI}
    I(A:B) = S(A) + S(B) - S(A\cup B),
\end{equation}
which will naturally reduce to the entanglement entropy if the bipartite system is taken to be the whole pure-state system. 

Another measure, known as the conditional mutual information (CMI), quantifies the correlations between two quantum systems $A$ and $B$, conditioned on a third system $E$. It is defined as 
\begin{equation}
    I(A:B|E) = S(A\cup E) + S(B\cup E) - S(A\cup B\cup E) - S(E).
\end{equation} 
It measures the information shared between $A$ and $B$ when $E$ is known. CMI is non-negative due to the strong subadditivity of the entanglement entropy, reflecting its role in characterizing quantum and classical correlations in multipartite systems. 

In holography, the Ryu-Takayanagi (RT) formula computes {relates} the entanglement entropy to the area of the minimal bulk surface homologous to the boundary entangling region. For CMI, this involves calculating entanglement entropies for boundary regions $A\cup E$, $B\cup E$, $A\cup B\cup E$, and $E$, each via their respective minimal surfaces. The combination of these areas yields the holographic CMI, which probes multipartite entanglement structures in the bulk \cite{Ju:2024hba,Ju:2024kuc}.

We are motivated to employ CMI to quantify the bipartite quantum entanglement in holography considering its advantages over mutual information. Briefly speaking, holographic CMI provides a phase-transition-free genuine measure of entanglement. For one thing, the mutual information $I(A:B)$ for boundary subregions $A$ and $B$, using RT prescription, exhibits a first-order phase transition when the minimal surface switches between the connected and disconnected configurations as $A$ and $B$ separate. 
{The phase transition of the holographic mutual information as the entangling regions separate represents a drop in correlations from $O(N^2)$ to $O(1)$ in the large-$N$ limit. Besides, the lack of correlations at large separation indicates the need to include bulk quantum corrections of $O(1)$.}
{In contrast, CMI does not possess this $O(N^2)$ order phase transition by conditioning on $E$}, ensuring a smooth behavior. For another thing, {at leading order,} holographic MI fails to capture correlations between distant $A$ and $B$ due to the existence of multipartite entanglement with other subsystems when the disconnected surfaces dominate. In contrast, CMI helps detect the non-vanishing entanglement between $A$ and $B$ by leveraging the conditioning on $E$. These advantages render CMI an indispensable tool for accurately quantifying the entanglement in holographic studies.

For infinitesimal subregions $A$ and $B$ separated at a distance $l$, we have \cite{Ju:2024xcn,Vidal:2014aal}
\begin{equation}
    I(A:B|E) = -\frac{d^2S}{dl^2},
\end{equation}
where $E$ is the interval between $A$ and $B$ with length $l$.

Direct computation of the entanglement measures poses significant challenges, with limited analytically solvable instances. The gauge/gravity duality, exemplified by the AdS/CFT correspondence, serves as a useful framework for investigating the entanglement phenomenon.
Notably, the AdS/CFT correspondence stands out as one of the most extensively explored instances among various versions of holographic dualities. In a broader sense, the holographic principle posits that physics in boundary field theories can be encapsulated in gravitational theories within the bulk spacetime, and vice versa.

A brief holographic prescription for the entanglement entropy goes as follows. {Generally, the HRT surface $\varepsilon_A$ can be found as the extremal bulk surface homologous to the boundary region $A$, with the entanglement entropy evaluated holographically at leading order as \cite{Ryu:2006bv}:
\begin{equation}\label{HEE}
    S(A)=\frac{\text{Area}(\varepsilon_A)}{4G_N},
\end{equation}
that is, $S(A)$ is proportional to the area of $\varepsilon_A$. A shortcut to fix this surface is the use of the max-min method \cite{Wall:2012uf}.} In static spacetimes, the method can be simplified as the RT prescription, i.e. searching for the minimal surface on the fixed spacelike time slice.



Here we present the detailed result of the holographic entanglement entropy for a general translationally symmetric background geometry. We start from the following asymptotic AdS$_4$ spacetime with the metric in Poincare coordinates 
\begin{equation}\label{metric_AdS4}
    ds^2=\frac{L^2}{z^2}[-f(z)dt^2+\frac{dz^2}{f(z)}+h(z)(dx^2+dy^2)],
\end{equation}
where $L$ denotes the AdS radius and the boundary is at {$z=0$}.

{The entanglement entropy corresponding to an effectively one-dimensional boundary strip region can be evaluated in this coordinate system as} \cite{Ahn:2024jkk}
\begin{equation}\label{HEE_AdS4}
    S=\frac{L^2\Omega}{2G_N}\int_{\epsilon}^{z_*} dz \frac{1}{z^2}\sqrt{\frac{h(z)}{1-\frac{z^4 h(z_*)^2}{z_*^4 h(z)^2}}}\frac{1}{\sqrt{f(z)}},
\end{equation}
where $\epsilon$ is the UV cutoff and the transitioning radial point $z_*$ is determined by the boundary length scale $l$
\begin{equation}\label{lzstar}
    l=2\int_0^{z_*}dz \frac{1}{\sqrt{\frac{h(z)^2 z_*^4}{h(z_*)^2 z^4}-1}}\frac{1}{\sqrt{h(z)f(z)}}.
\end{equation}
In general, there is no analytic expression for $S(l)$. Instead, combining the two relations above would yield a simple relation of $S$, $l$, and $z_*$:
\begin{equation}\label{mainformula}
    \frac{dS}{dl}=\frac{dz_*}{dl}\frac{dS}{dz_*}=\frac{L^2\Omega}{4G_N}\frac{h(z_*)}{z_*^2}.
\end{equation}
The negative value of the derivative of LHS simply amounts to the CMI of two parallel boundary strips $A$ and $B$ with their lengths being infinitesimal and the distance in between being $l$ \cite{Ju:2024xcn,Vidal:2014aal}, i.e. 
\begin{equation}\label{cmi}
   I(A,B|E) = -\frac{d^2S}{dl^2},
\end{equation}
where $E$ is the subregion between $A$ and $B$ with length $l$.

This indicates that for a given geometry {in which the relation $z_*(l)$ is obtained and substituted into the expression for CMI, it is plausible that the CMI has a local dependence on the bulk metric at $z=z_*$.} However, as we have already explained in section \ref{1}, {the extraction of this seemingly local correspondence} requires us to know the dependence of $z_*$ on $l$ a prior, which also depends on the geometry in the outer UV region. This is because when we take the derivative of $l$ on the RHS of \eqref{mainformula}, there will be an extra contribution of $d z_*/d l$. In a general background geometry, the relation between $l$ and $z_*$ in \eqref{lzstar} is an integration function that involves geometry at all $z\leq z_*$, making the CMI an integration function which cannot be locally determined by the geometry at $z_*$. 

Inspired by the argument that the boundary CMI/bulk radial geometry correspondence should be gauge dependent and to solve the problem of not knowing the relation $z_*(l)$ in advance, we use the function $z_*(l)$ as the specification of the gauge choice. {In fact, we can use an arbitrary smooth function} $z_*(l)$ as the prior criterion that fixes the gauge choice of the coordinate system. In the next subsection, we will show the explicit CMI/bulk radial geometry correspondence in the simplest gauge where $z_*(l)=l$. This gauge choice can be implemented by transforming the old metric into a new coordinate system corresponding to this gauge choice. 

\subsubsection{Specifying the gauge choice}\label{The specific gauge choice}

In this subsection, we show the metric {in a specific gauge}, which we choose to be $z_*(l)=l$ for simplicity. Starting from the original coordinate system in \eqref{metric_AdS4} {and the relation $z_*(l)$ which can be evaluated numerically} from \eqref{lzstar}, we perform a coordinate transformation of the radial coordinate $z\rightarrow\rho$ so that
\begin{equation}\label{ztorho}
    l=c \rho_*=2\int_0^{z_*}dz \frac{1}{\sqrt{\frac{h(z)^2 z_*^4}{h(z_*)^2 z^4}-1}}\frac{1}{\sqrt{h(z)f(z)}},
\end{equation}
where $c$ is an arbitrary {nonzero} constant.
The second equality above defines the radial coordinate transformation as a one-to-one correspondence between $\rho$ and $z$ indicated by the relation of $\rho_*(z_*)$ at arbitrary value of $z_*$  \footnote{Note that when there is an entanglement shadow in the bulk spacetime where all $z_*$ is in the exterior region, this reconstruction would not work in that entanglement shadow.}. This is in general a nonlinear coordinate transformation. 
The first identity in (\ref{ztorho}) establishes the new gauge choice $\rho_*=l$ {such that it defines} the linear relation of the boundary length scale $l$ and the special radial position $\rho_*$ in the new coordinate system. Under this coordinate transformation, the relation of the boundary scale $l$ and the bulk radial scale $z_*$ in \eqref{mainformula} now becomes linear in the new $\rho_*$ coordinate system. We will set the constant $c\equiv 1$ in the following sections for convenience.

The bulk metric in \eqref{metric_AdS4} now takes the following form in the $\rho$-coordinate system
\begin{equation}\label{metric_new}
    ds^2=\frac{L^2}{\rho^2}[-\tilde{F}(\rho)dt^2+\frac{d\rho^2}{F(\rho)}+H(\rho)(dx^2+dy^2)],
\end{equation}
where $\tilde{F}$ and $F$ are two different functions in general. In this way, the induced metric on the Cauchy slice takes a similar form as in the $z$-coordinate system while all the effects of the radial transformation has been absorbed in the new metric functions $\tilde{F}$, $F$, and $H$. Therefore, all the formulae obtained in $z$-coordinate system are still applicable in $\rho$-coordinate system. 

Comparing \eqref{metric_AdS4} and \eqref{metric_new}, it is easy to check that the metric functions in $z$- and $\rho$-coordinates are related by
\begin{equation}\label{transformation_metric_functions}
\begin{split}
\begin{aligned}
    H(\rho)&=\frac{\rho^2}{z(\rho)^2}h(z(\rho)),\\
    F(\rho)&=\frac{z(\rho)^2}{z'(\rho)^2\rho^2}f(z(\rho)),
\end{aligned}
\end{split}
\end{equation}
where $h$ and $f$ in $z$-coordinate system are transformed to be $H$ and $F$ in $\rho$-coordinate system, respectively. 

Therefore, the entanglement entropy can be computed from \eqref{mainformula} in the $\rho$-coordinate system
\begin{equation}\label{mainformula_new}
    \frac{dS}{dl}=\frac{dS}{d\rho_*}=\frac{L^2\Omega}{4G_N}\frac{H(\rho_*)}{\rho_*^2}.
\end{equation}
Note that the entanglement entropy $S(l)$ is invariant under the radial coordinate transformation, which will be manifest in specific examples.

The crucial point is now in the new $\rho$-coordinate system, $l= \rho_*$, and therefore the CMI of infinitesimal subregions $A$ and $B$ under the condition of the region $E$ in between with length $l$ is
\begin{equation}\label{cmirho}
   I(A,B|E) = -\frac{d^2S}{dl^2}=-\frac{d}{ d\rho_*}\bigg(\frac{L^2\Omega}{4G_N}\frac{H(\rho_*)}{\rho_*^2}\bigg),
\end{equation}
which is a local function of $\rho_*$. This confirms that in this $\rho$-coordinate system, the CMI at a distance $l$ is fully determined by bulk geometry at $\rho=l$. Consequently, in this particular gauge choice of coordinate system, there is indeed an exact correspondence between the bulk geometry at a radial scale $l$ and the boundary real-space entanglement structure at a distance $l$. In fact, the fixed relation of $l=\rho_*$ imposes a constraint on the metric in the $\rho$- coordinate system and it has already encoded this constraint in the requirement of $l=\rho_*$. Therefore, the CMI at a real-space length $l$ is determined by the bulk geometry at $\rho_* = l$ in the special gauge choice of the coordinate system. 

We emphasize here that in each coordinate system, we would always have a local function of $z_*(l)$ indicating a local one-to-one correspondence {between} the radial geometry and boundary CMI behavior. However, this function varies in different coordinate systems. This {difference} reflects the fact that the radial position/boundary real-space scale correspondence $z_*(l)$ involves the UV geometry outside $z_*$, and is inherently nonlocal. Here we fix this correspondence $z_*(l)$ {as an a prior condition} so that in all geometries with appropriate gauge choices we require the same $z_*(l)$. This makes it appear as though there is a local one-to-one correspondence of $z_*(l)$, seemingly independent of the geometries. However, this does not change the fact that this correspondence is nonlocal. What we are doing here is to highlight the {\it nonlocality of the bulk radial geometry/boundary real-space entanglement correspondence}, {while demonstrating how this gauge choice can be employed to manifest a seemingly local correspondence}. This procedure will be helpful when we we reverse the process to reconstruct the bulk geometry from boundary CMI data.

In the following subsections, we perform this radial coordinate transformation and show with specific examples that the CMI is fully determined by the bulk geometry at a corresponding radial scale in this specific gauge choice.

\subsection{Explicit examples}

To substantiate the discussion in the previous subsection, we provide several examples in this subsection, including the pure AdS spacetime, black holes in three and four dimensions and a special geometry flowing from AdS$_2$ at the horizon to AdS$_4$ at the boundary that involves different scaling behaviors at different radial scales. We will continue to pick the $z_*=l$ gauge in the following sections.

\subsubsection{Pure AdS} 

We start with the pure AdS background, the simplest case. In four dimensions, the metric takes the form of \eqref{metric_AdS4} with $f(z)=h(z)=1$. 
The entanlgement entropy in \eqref{HEE_AdS4} and the boundary length in \eqref{lzstar} have analytic expressions:
\begin{equation}\label{AdS4_EE}
\begin{split}
\begin{aligned}
    S(l)&= \frac{L^2\Omega}{2G_N}\Big[ \frac{1}{\epsilon}-\frac{2\pi}{l}\Big( \frac{\Gamma(\frac{3}{4})}{\Gamma(\frac{1}{4})}\Big)^2 \Big],\\
    l(z_*)&= \frac{2\sqrt{\pi}\Gamma(\frac{3}{4})}{\Gamma(\frac{1}{4})}z_*.
\end{aligned}
\end{split}
\end{equation}
The first equation in \eqref{AdS4_EE} is exactly the scaling law of the entanglement entropy that we aim to construct. Note that this relation can be equivalently obtained using \eqref{mainformula}. Thus the relationship of the boundary length $l$ and the bulk radial position $z_*$ is already linear in $z$-coordinate system without any additional radial coordinate transformation in the bulk. Therefore, we need not perform the $z\rightarrow\rho$ transformation in this special AdS case as stated in the general description in section \ref{The specific gauge choice}.

Whereas, in a general case, the relationship of $l$ and $z_*$ would be a different function where all $z\leq z_*$ geometry plays a role, and therefore the coordinate transformation $z\rightarrow\rho$ would be necessary to perform as a gauge choice.

\subsubsection{BTZ black hole}
The metric for the three dimensional BTZ black hole takes the following form  
\begin{equation}
 ds^2  = \frac {L^2}{z^2}\left[-(1-Mz^2)dt^2+\frac1{1-Mz^2}dz^2+dx^2\right],  
\end{equation} {where $M$ is the mass of the black hole.}

The entanglement entropy can be evaluated in this coordinate system as
\begin{equation}
S(l)  = \frac{L}{2G}\ln\left(\frac{2}{\sqrt{M}\varepsilon}\sinh\frac{\sqrt{M}l}{2}\right)
\end{equation}

Similar to (\ref{lzstar}), the relationship between $l$ and $z_*$ under the BTZ black hole can be calculated.
\begin{equation}
l= 2\int_0^{z_*}\frac{z}{\sqrt{z_*^2-z^2}}\frac1{\sqrt{1-Mz^2}}dz  = \frac{2}{\sqrt{M}}\tanh^{-1}(\sqrt{M}z_*)
\end{equation}
This is a local function of $z_*$ which however does involve the metric at $z<z_*$ as the metric is known and has been integrated out.

For pure $AdS_{d+2}$ with $M=0$ it simplifies to the known result
\begin{equation}
    l=2\sqrt{\pi}\frac{\Gamma(\frac{d+1}{2d})}{\Gamma(\frac{1}{2d})}z_*,
\end{equation}

and when d=1, $l=2z_*$.
For the three dimensional BTZ black hole, consider the coordinate transformation $z=z(\rho)$ such that $l=\rho_*$. Consequently, the specific form of the coordinate transformation can be obtained as follows
\begin{equation}
    z=\frac{1}{\sqrt{M}}\tanh(\frac12\sqrt{M}\rho).
\end{equation}
The metric after coordinate transformation becomes

\begin{equation}
    ds^2  = \frac {L^2}{\rho^2}\left[-T(\rho)dt^2+\frac{1}{F(\rho)}d\rho^2+H(\rho)dx^2\right],
\end{equation}
where compared to the original metric, $F(\rho)$ and $H(\rho)$ become
\begin{equation}
    F(\rho) = \frac{z^2f(z)}{\rho^2z'(\rho)^2}  = \left(\frac{2\sinh(\frac12\sqrt{M}\rho)}{\sqrt{M}\rho}\right)^2,
\end{equation}
\begin{equation}
    H(\rho) = \frac{\rho^2}{z^2}h(z)  = \left(\frac{\sqrt{M}\rho}{\tanh{\frac12\sqrt{M}\rho}}\right)^2.
\end{equation} Therefore we have obtained the new coordinate system for the BTZ black hole in which the boundary real-space entanglement behavior indicated by CMI at the scale $l$ is determined by bulk geometry at radial position $\rho=l$.

\subsubsection{AdS$_4$ Schwarzschild black holes}\label{2.2.3}

\begin{figure}[!htbp]
    \centering
      \includegraphics[width=0.6\textwidth]{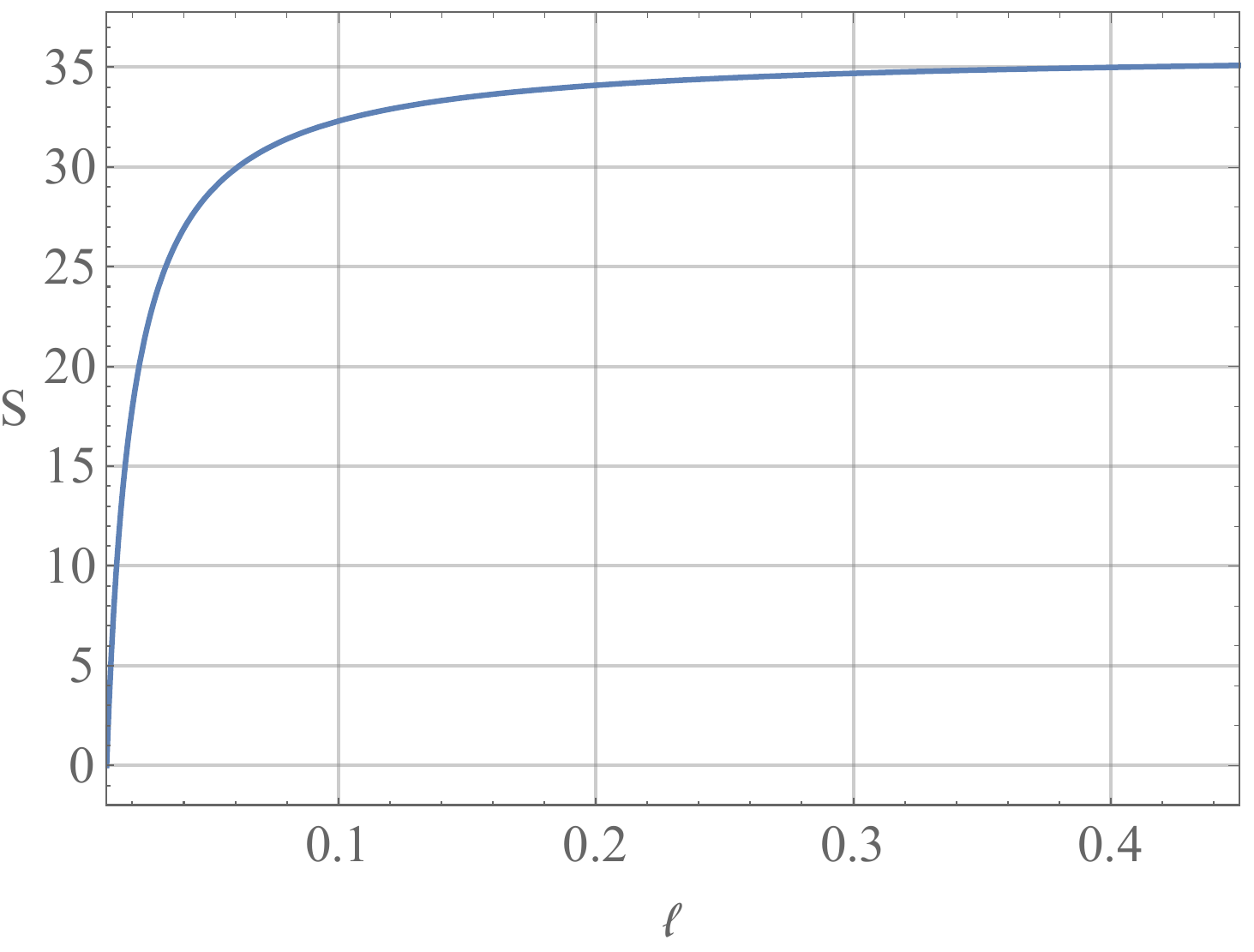}
    \caption{The entanglement entropy evaluated from the background metric in AdS black hole geometry {with the UV divergence removed}. Parameter setting: $L = 1$, $\Omega = 1$, $G_N=1$.}
    \label{fig_EE_AdSBH(1)}
\end{figure}

Next, we discuss the AdS$_4$ Schwarzschild black hole as our third example. The black hole metric takes the form of \eqref{metric_AdS4} with $f(z)=1-(\frac{z}{z_0})^3$ and $h(z)=1$. {For this case, the gauge choice $z_*=l$ has to be found  numerically as there is no analytic result when integrating the geometry in the UV region.} Insert $f(z)$ and $h(z)$ into \eqref{ztorho}, and we obtain the numerical relation of the old coordinate $z$ and the new coordinate $\rho$. With this numerical relation, we evaluate the scaling law of the entanglement entropy by plugging the inverse relation $z(\rho)$ into \eqref{mainformula}. The graphical result is shown in figure \ref{fig_EE_AdSBH(1)}. Note that the entanglement entropy is renormalized as $S(l)-S(\epsilon)$ where $\epsilon$ is the UV cutoff rendering the entropy difference finite.

\begin{figure}[!htbp]
    \centering
      \includegraphics[width=\textwidth]{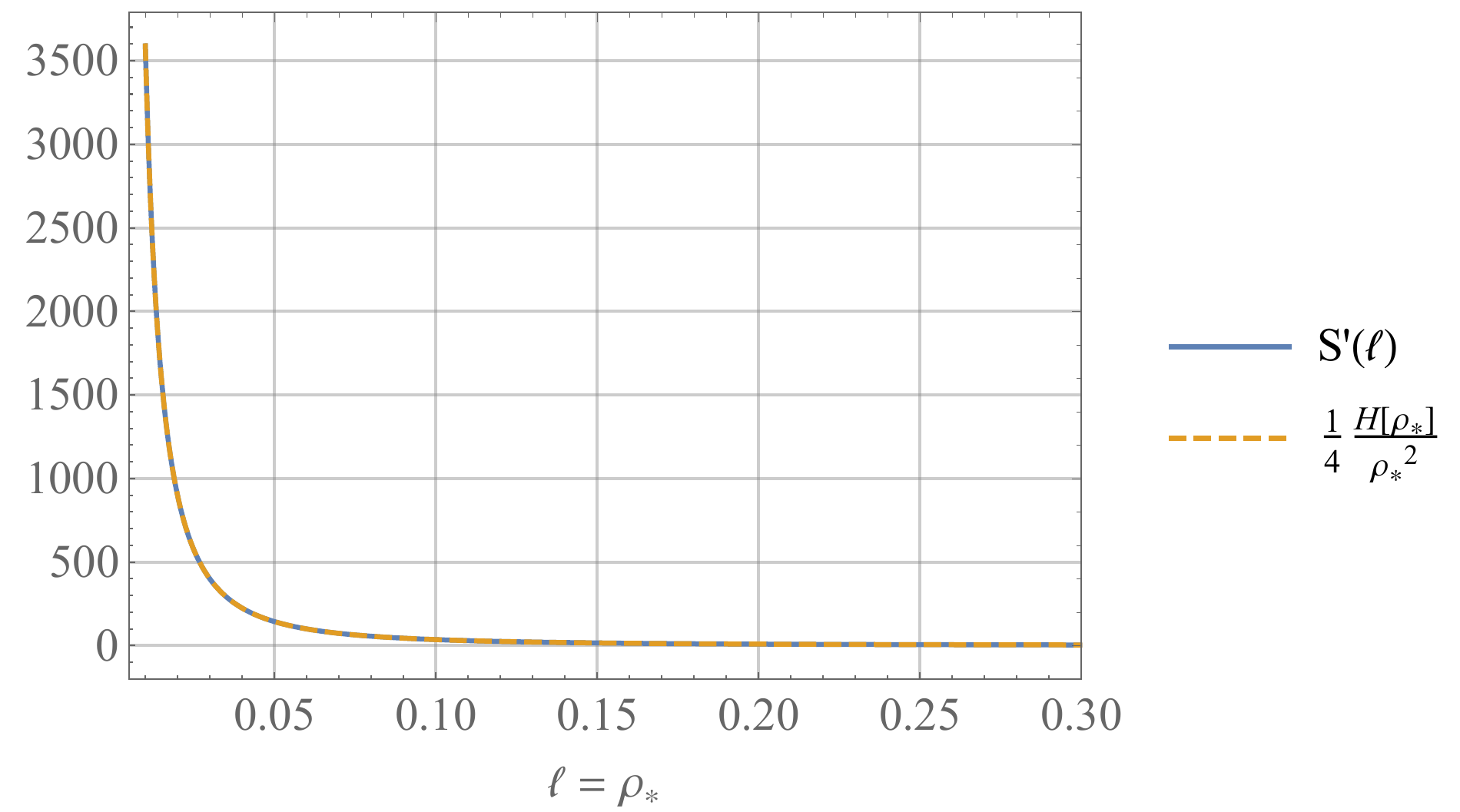}
    \caption{Comparison of the entanglement data and the bulk geometry data in AdS black hole geometry. The solid blue curve denotes the derivative of entanglement entropy $S'(l)$, computed as the LHS of \eqref{mainformula}. The dashed yellow curve denotes the RHS of \eqref{mainformula_new} with the identification of $l=\rho_*$. Parameter setting: $L = 1$, $\Omega = 1$, $G_N=1$. For simplicity and higher numerical precision, we have used the first-order derivative $S'(l)$, i.e. the LHS of \eqref{mainformula} instead of the second-order derivative $S''(l)$ which corresponds directly to the CMI.}
    \label{fig_EE_AdSBH(2)}
\end{figure}

With the relations of the metric functions in $z$- and $\rho$-coordinate systems given in \eqref{transformation_metric_functions}, we compare the entanglement entropy data (CMI data)  computed as the LHS of \eqref{mainformula} with the bulk geometry data at the radial position $\rho_*$ computed as the RHS of \eqref{mainformula_new}, as displayed in figure \ref{fig_EE_AdSBH(2)}. The match of the two curves indicates that the bulk geometry at $\rho_*$ determines the boundary CMI behavior at real-space scale $l$. 

\subsubsection{A flow geometry in Einstein-Maxwell-dilaton theory}\label{2.2.4}

In this example, we consider in Einstein-Maxwell-dilaton (EMd) theory a flow geometry exhibiting different scaling behaviors at different radial scales. The geometry flows from AdS$_2\times$R$_2$ in IR to AdS$_4$ in UV, possessing an intermediate hyperscaling violating geometry \cite{Bhattacharya:2012zu,Kundu:2012jn}. A brief framework is summarized as follows.

EMd theory is described by the Lagrangian 
\begin{equation}
    \textit{}{L} = R-2(\partial\phi)^2+f(\phi)F_{\mu\nu}F^{\mu\nu}-V(\phi),
\end{equation}
where $\phi$ denotes the dilatonic scalar field, $F_{\mu\nu}$ denotes the field strength of the Maxwell gauge field $A_\mu$, and $V$ denotes the self-interacting potential of the scalar field. Using functional derivatives, one can derive the corresponding equations of motion for the gravitational field, scalar field, and gauge field respectively as
\begin{equation}
\begin{split}
\begin{aligned}
    R_{\mu\nu}+\frac{1}{2}(V(\phi)-R)g_{\mu\nu} &= 2\partial_{\mu}\phi\partial_{\nu}\phi-g_{\mu\nu}\partial_{\rho}\phi\partial^{\rho}\phi
    \\
    &-2f(\phi)(F_{\mu\rho}F^{\rho}_{\nu}-\frac{1}{4}g_{\mu\nu}F_{\rho\sigma}F^{\sigma\rho}),\\
    4\nabla_\mu\partial^\mu\phi &= \partial_\phi f(\phi)F_{\mu\nu}F^{\mu\nu} + \partial_\phi V(\phi),\\
    \nabla_\mu(f(\phi)F^{\mu\nu}) &= 0.
\end{aligned}
\end{split}
\end{equation}

The hyperscaling violating geometry can be found as a typical analytic solution to this theory (in arbitrary dimensions):
\begin{equation}
    ds^2 = r^{-\frac{2(d-\theta)}{d}} (-r^{-2(z-1)}dt^2+dr^2+dx^2+dy^2+...),
\end{equation}
which is parametrized by the hyperscaling violation exponent $\theta$ and the dynamical critical exponent $z$. One can check that this solution reduces to AdS geometry in the case of $(z,\theta)=(1,0)$. 

Generally, one can use the following isotropic ansatz for the metric:
\begin{equation}\label{metric_HV(1)}
    ds^2 = -a(r)^2dt^2+\frac{dr^2}{a(r)^2}+b(r)^2(dx^2+dy^2+...),
\end{equation}
which we will adopt in 3+1 dimensions hereinafter.

\begin{figure}[!htbp]
    \centering
    \begin{subfigure}[b]{0.45\textwidth}
      \includegraphics[width=\textwidth]{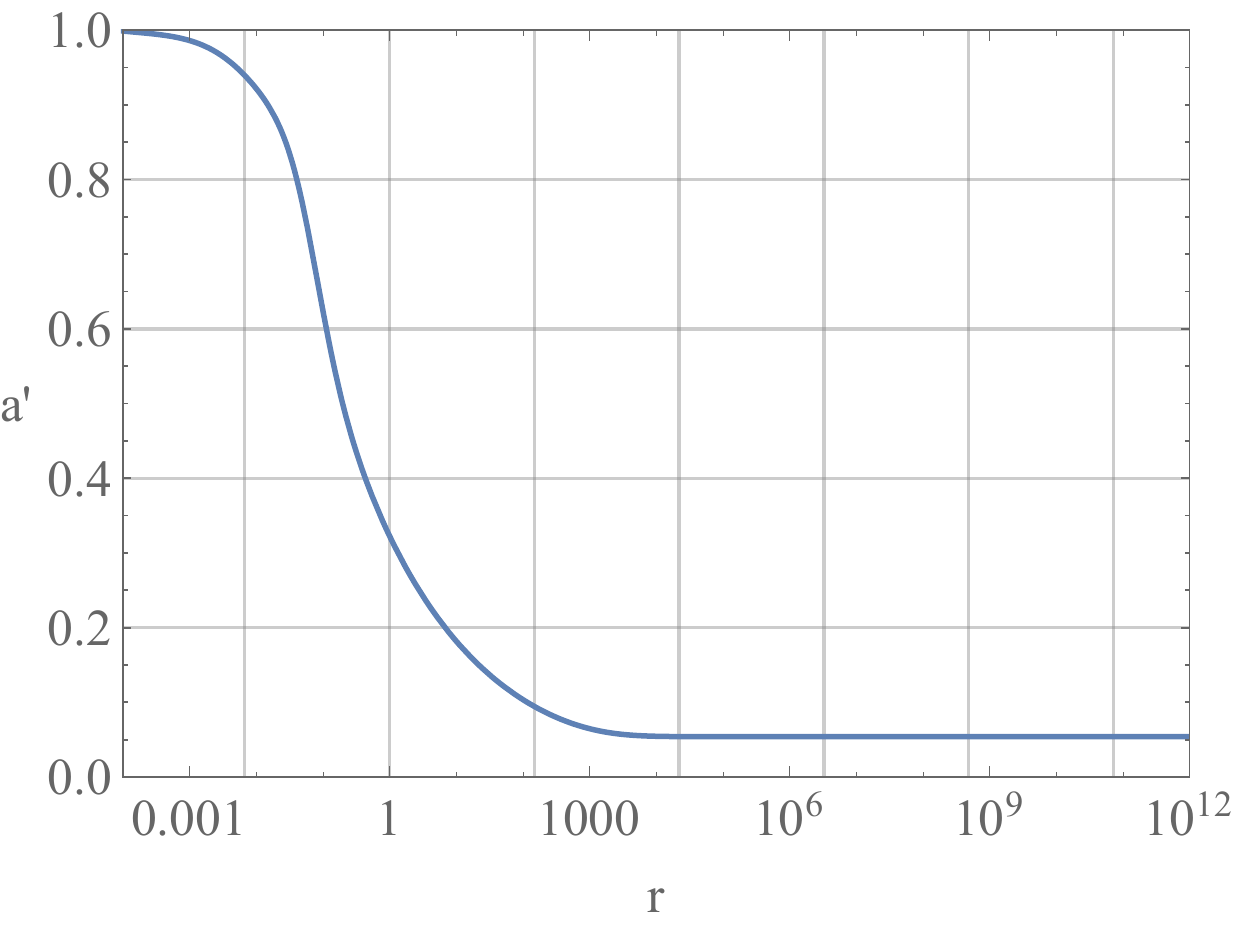}
      \caption{$a'(r)$.}
    \end{subfigure}
    \begin{subfigure}[b]{0.45\textwidth}
      \includegraphics[width=\textwidth]{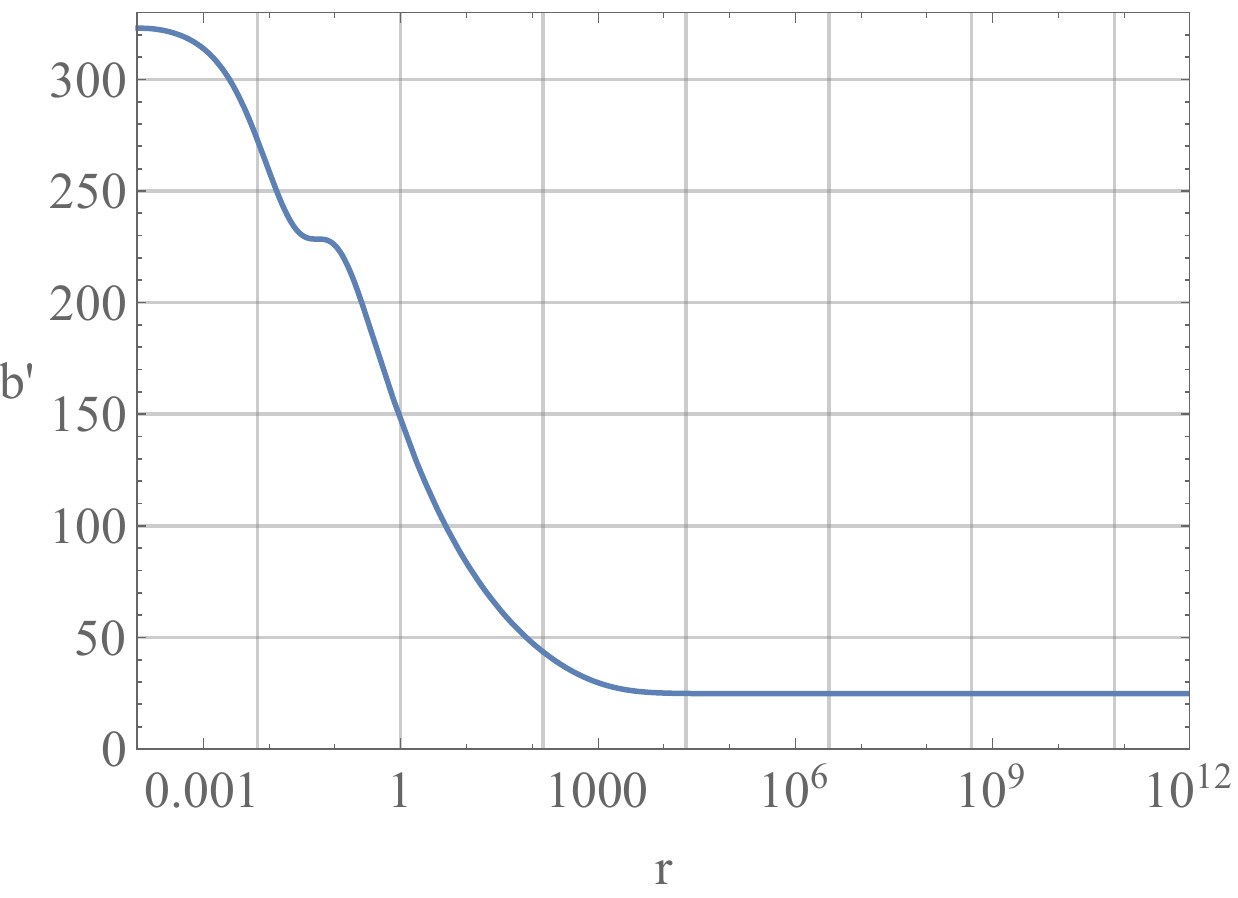}
      \caption{$b'(r)$.}
    \end{subfigure}
    \begin{subfigure}[b]{0.45\textwidth}
      \includegraphics[width=\textwidth]{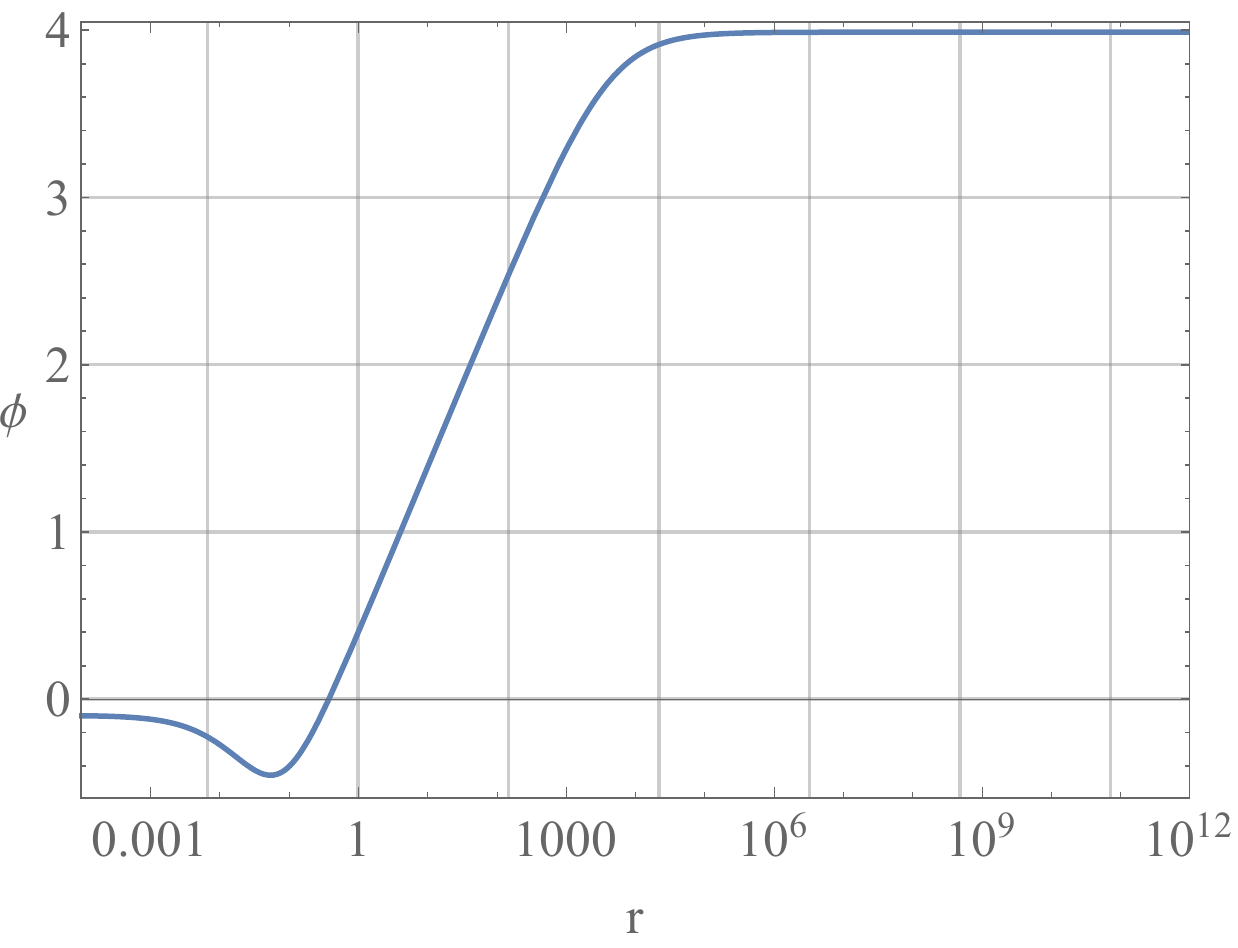}
      \caption{$\phi(r)$.}
    \end{subfigure}
    \caption{The numerical solutions of the metric and scalar functions $a(r)$, $b(r)$, and $\phi(r)$ in the flow geometry. The prime sign denotes the radial derivative.}
    \label{fig_numerical_bkgrd_HV}
\end{figure}

Following \cite{Bhattacharya:2012zu}, we evaluate by using the general metric ansatz above the bulk solution with a running flow from IR to UV, in which the IR geometry is AdS$_2\times$R$_2$ and the UV geometry is asymptotic AdS$_4$. For the purpose of numerical computation, the gauge field is assumed to be a constant magnetic field $F = Q_m dx \land dy$, and the scalar functionals $f$ and $V$ are assumed to take the form of
\begin{equation}
\begin{split}
\begin{aligned}
    f(\phi(r)) &= e^{2\alpha\phi(r)} + \xi_1 + \xi_2 e^{-2\alpha\phi(r)},\\
    V(\phi(r)) &= -V_0(e^{-\eta\phi(r)} + c_1 e^{\eta\phi(r)}),
\end{aligned}
\end{split}
\end{equation}
with the IR behavior of the fields constrained by
\begin{equation}
\begin{split}
\begin{aligned}
    a(r\rightarrow 0) &= r(1+d_1 r^\nu+\cdots),\\
    b(r\rightarrow 0) &= b_H(1+d_2 r^\nu+\cdots),\\
    \phi(r\rightarrow 0) &= \phi_H(1+d_3 r^\nu+\cdots).
\end{aligned}
\end{split}
\end{equation}
The numerical solutions are displayed in figure \ref{fig_numerical_bkgrd_HV}, with the parameters fixed as
\begin{equation}
\begin{split}
\begin{aligned}
    \alpha &= \sqrt{3},\;\xi_1 = 0,\;\xi_2 = 1,\\
    V_0 &= 3\times 10^4,\;c_1 = 10^{-4},\;\eta = \frac{2}{\sqrt{3}},\\
    \nu &= 1,\;d_1 = -5.558,\;d_2 = -2.99 d_1,\;d_3 = 5.19 d_1,\\
    \phi_H &= -0.1,\;Q_m = 2,\; b_H = 13.74\sqrt{Q_m},
\end{aligned}
\end{split}
\end{equation}
along with the IR cutoff $r_{IR} = 10^{-4}$ and UV cutoff $r_{UV} = 10^{12}$.

To fit in with the metric ansatz in \eqref{metric_AdS4}, we identify the metric functions using the transformation of the radial coordinates $z = 1/r$: 
\begin{equation}
    f(z)=z^2a(\frac{1}{z})^2,\;\;h(z)=z^2b(\frac{1}{z})^2,
\end{equation}
from which we obtain data of $f(z)$ and $h(z)$\footnote{The numerical computation would be difficult due to the numerical solution of the background metric and the numerical integration (in \eqref{ztorho}, for instance). To deal with this technical difficulty, we fit the metric functions numerically at a sufficiently high order.}.

\begin{figure}[!htbp]
    \centering
    \includegraphics[width=0.6\textwidth]{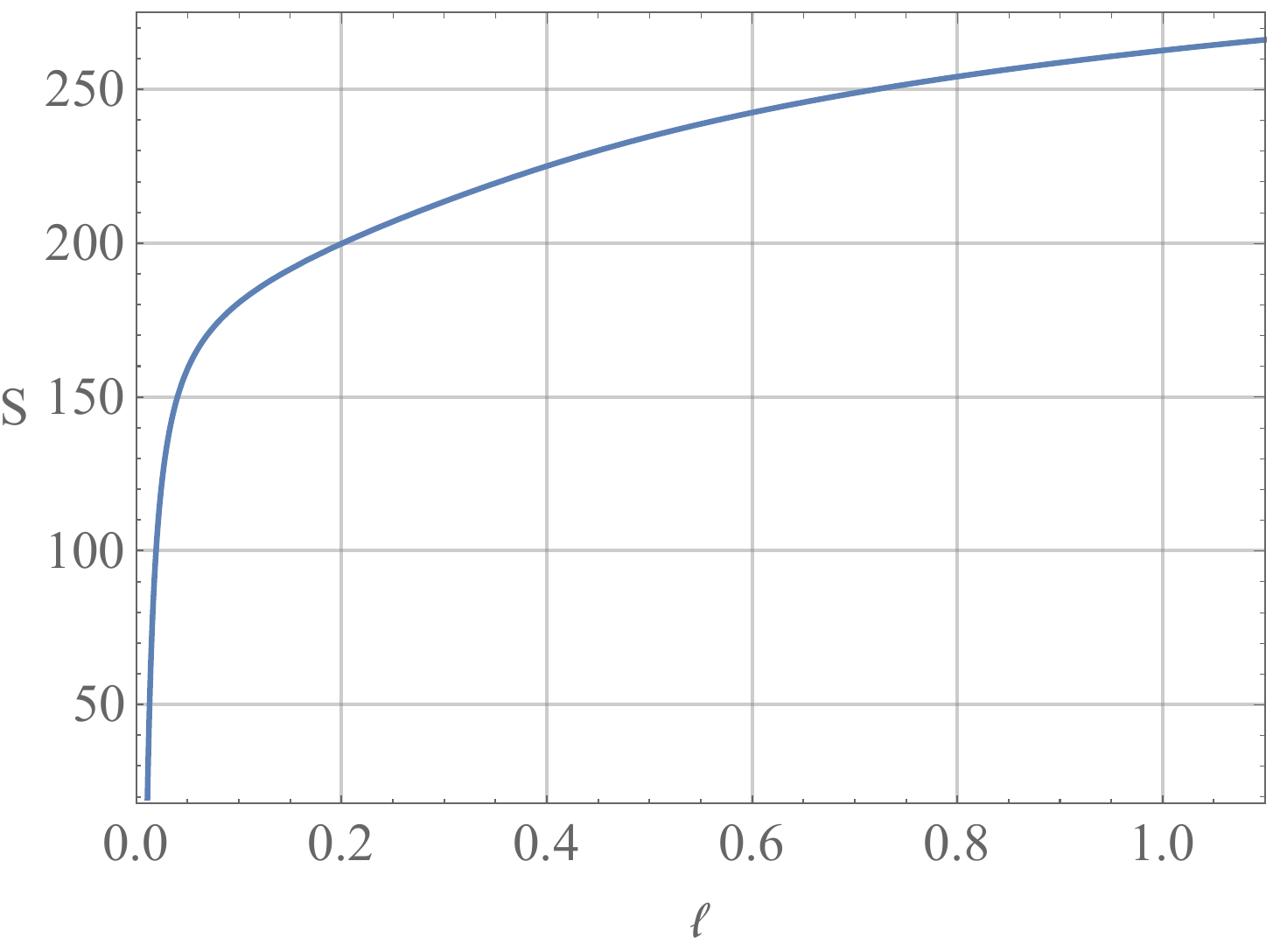}
    \caption{The entanglement entropy evaluated from the background metric in the flow geometry {with the UV divergence removed}. Parameter setting: $L = 1$, $\Omega = 1$, $G_N=1$.}
    \label{fig_EE_HV(1)}
\end{figure}

Insert the numerical solutions of $f(z)$ and $h(z)$ into \eqref{ztorho}, and we obtain the numerical relation of the old coordinate $z$ and the new coordinate $\rho$ with the new coordinate which is defined by the gauge of $\rho_*=l$. We evaluate the scaling law of the entanglement entropy by plugging the inverse relation $z(\rho)$ into \eqref{mainformula}, as shown in figure \ref{fig_EE_HV(1)}. Note that the cutoff term $S(\epsilon)$ has been subtracted in the entanglement entropy.

\begin{figure}[!htbp]
    \centering
    \begin{subfigure}[b]{0.45\textwidth}
      \includegraphics[width=\textwidth]{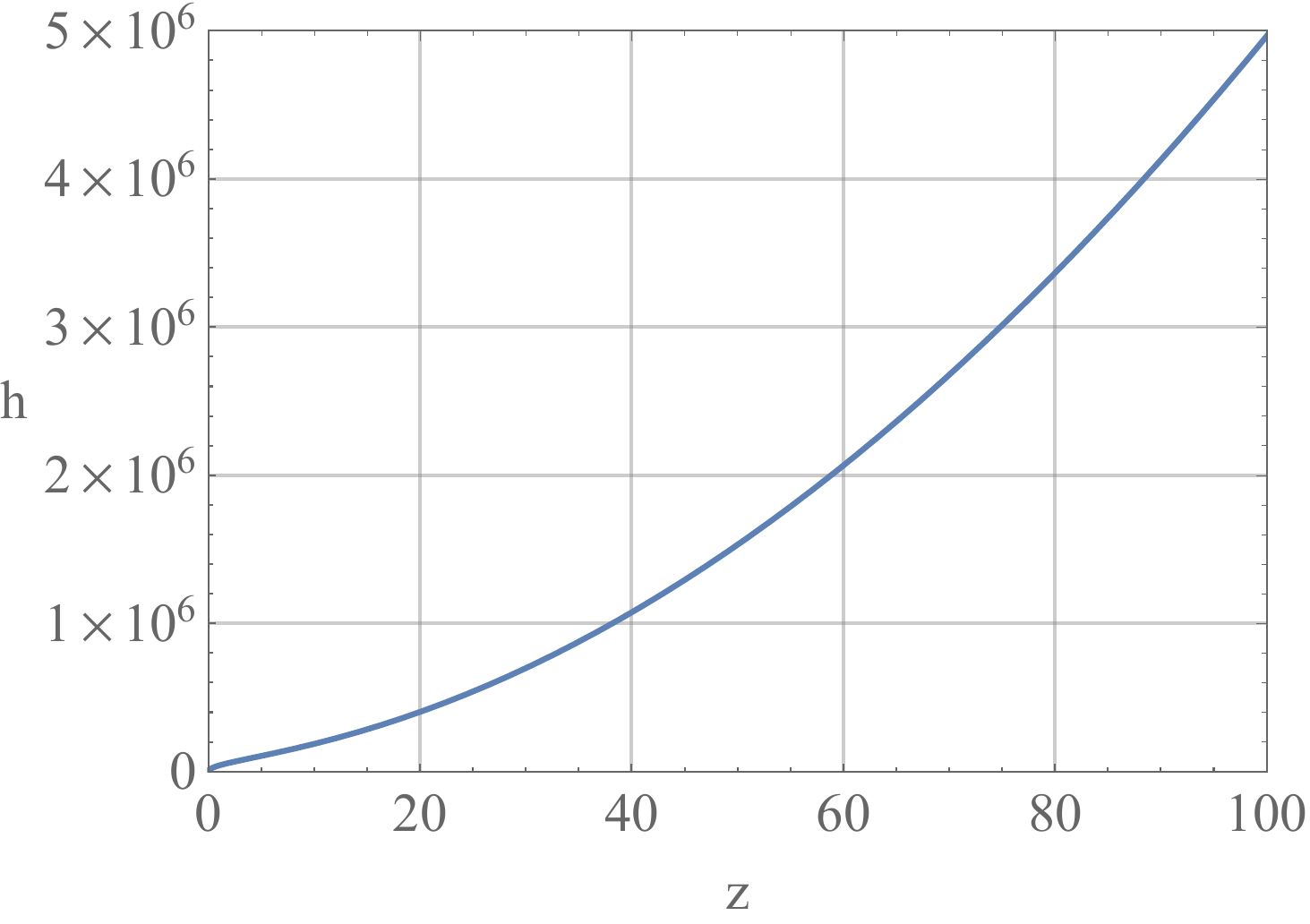}
      \caption{$h(z)$.}
    \end{subfigure}
    \begin{subfigure}[b]{0.45\textwidth}
      \includegraphics[width=\textwidth]{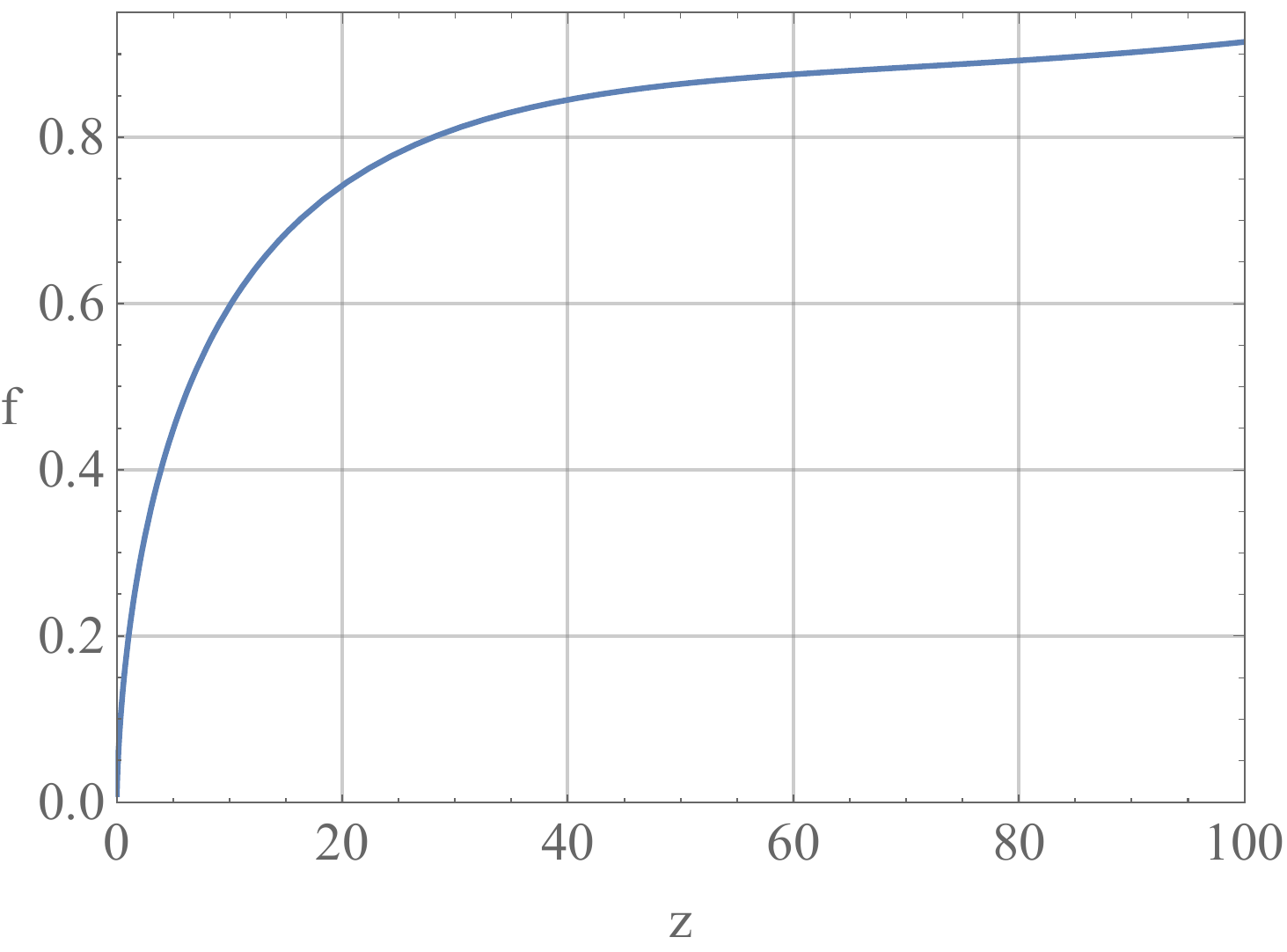}
      \caption{$f(z)$.}
    \end{subfigure}
    \begin{subfigure}[b]{0.45\textwidth}
      \includegraphics[width=\textwidth]{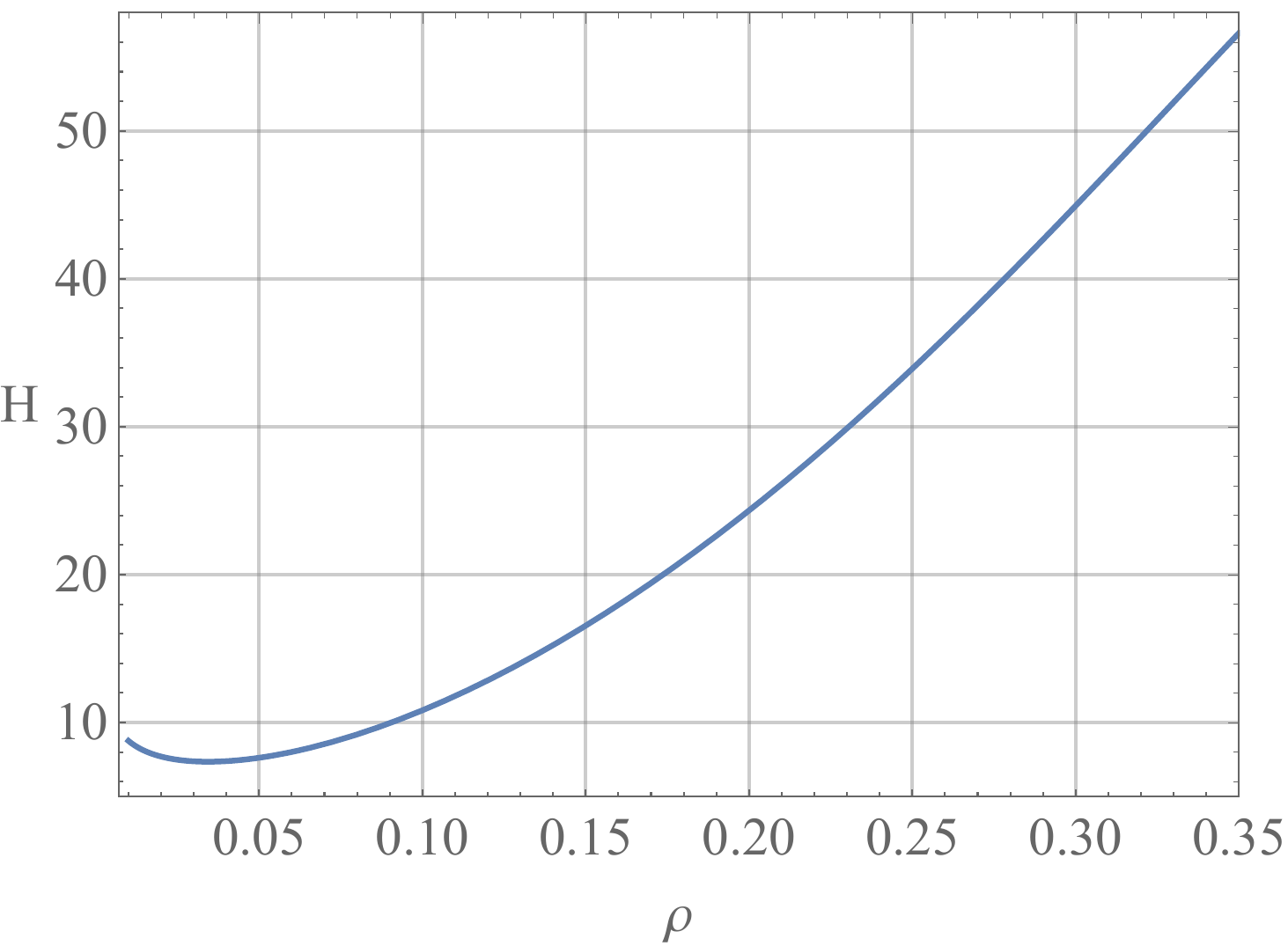}
      \caption{$H(\rho)$.}
    \end{subfigure}
    \begin{subfigure}[b]{0.45\textwidth}
      \includegraphics[width=\textwidth]{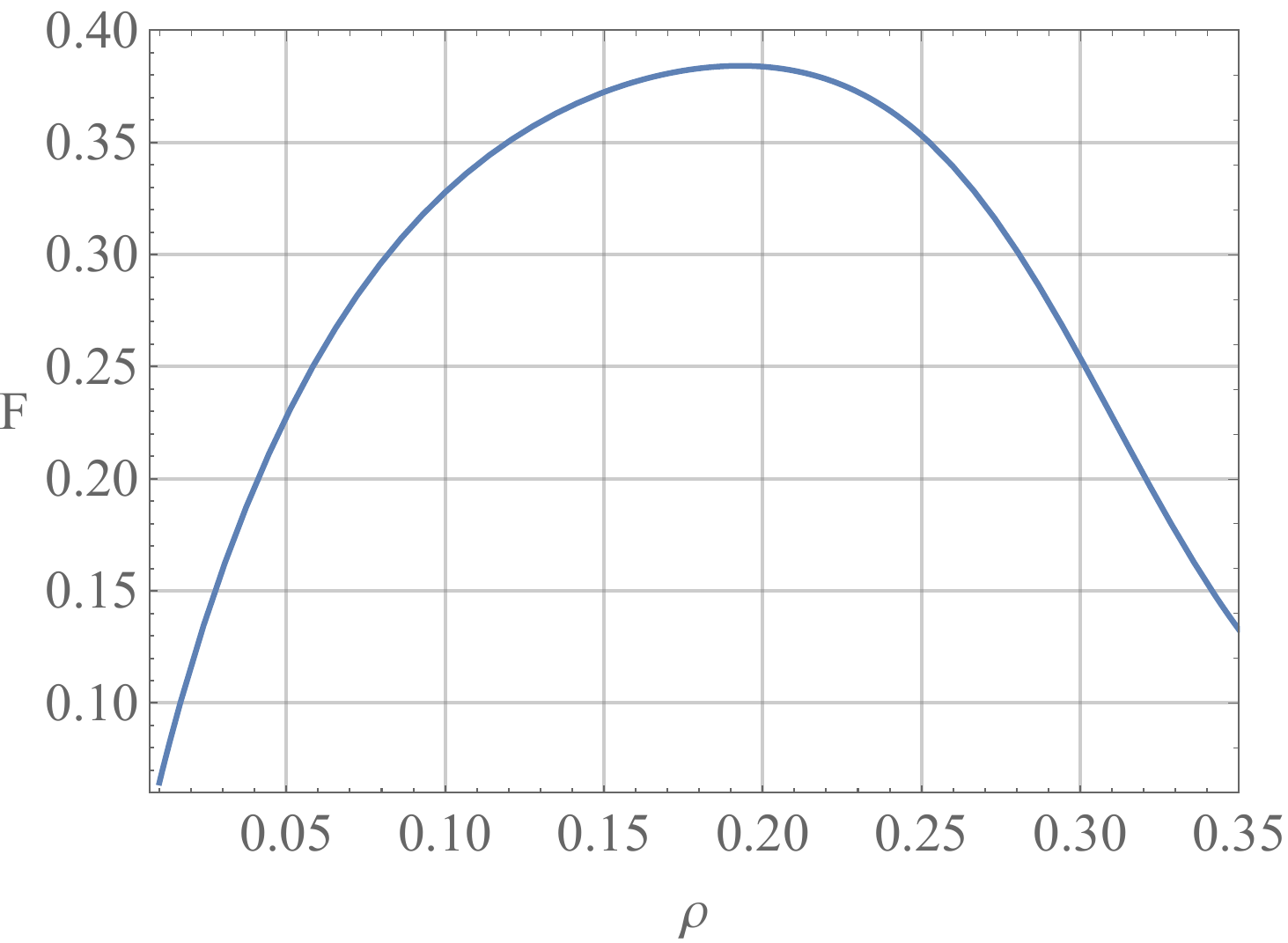}
      \caption{$F(\rho)$.}
    \end{subfigure}
    \caption{Metric functions in $z$- and $\rho$-coordinate systems in the flow geometry. With the radial coordinate transformation $z\rightarrow\rho$, metric functions $h$ and $f$ are transformed to be $H$ and $F$, respectively.}
    \label{fig_metric_functions_HV}
\end{figure}

With the general transformation rule of the metric functions in $z$- and $\rho$-coordinate systems given in \eqref{transformation_metric_functions}, numerical results of the metric functions $h(z)$, $f(z)$, $H(\rho)$, and $F(\rho)$ in this flow geometry are computed and displayed in figure \ref{fig_metric_functions_HV}. 

\begin{figure}[!htbp]
    \centering
    \includegraphics[width=\textwidth]{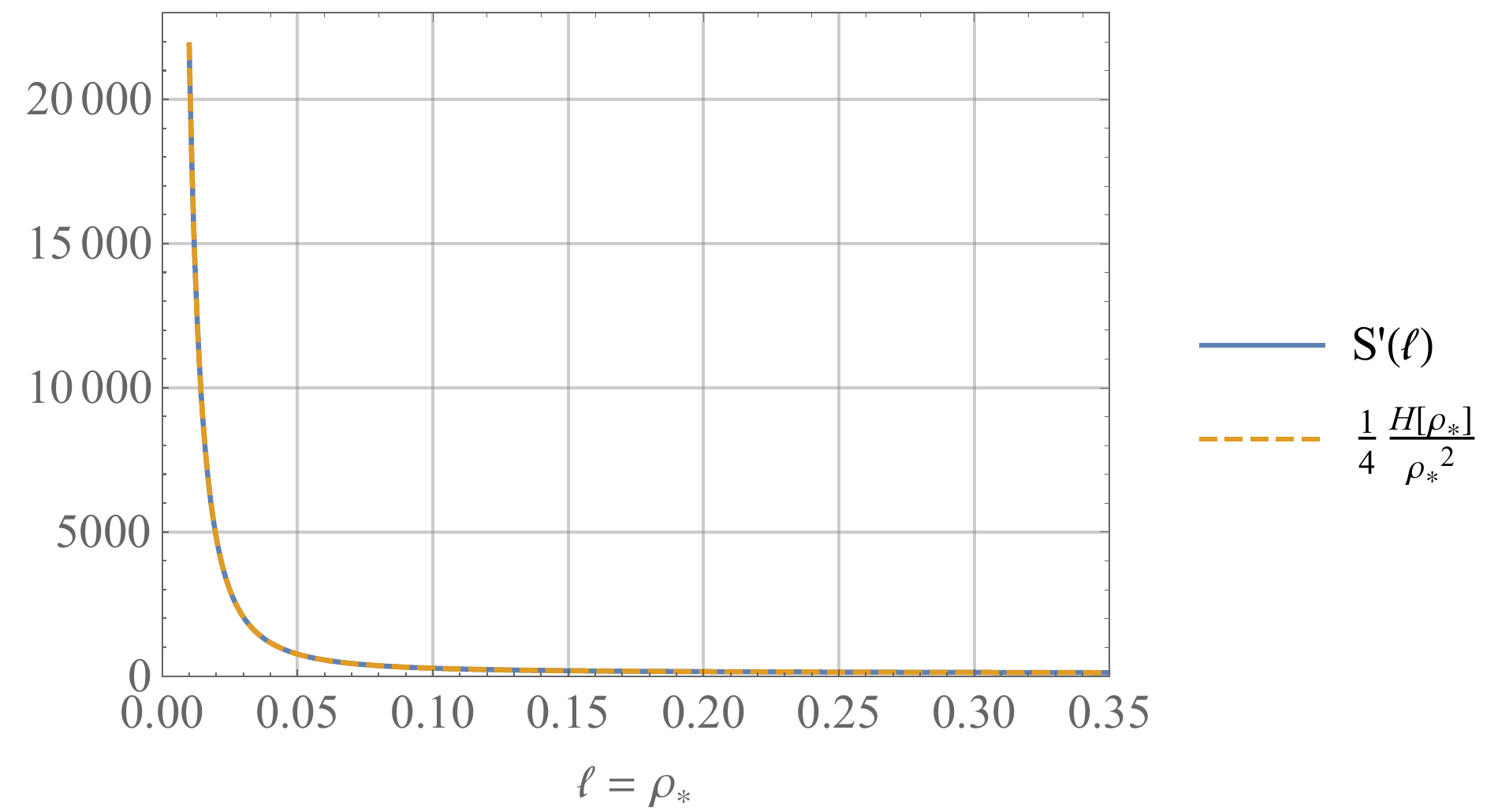}
    \caption{Comparison of the entanglement data and the bulk geometry data in the flow  geometry. The solid blue curve denotes the derivative of entanglement entropy $S'(l)$, computed as the LHS of \eqref{mainformula}. The dashed yellow curve denotes the RHS of \eqref{mainformula_new} with the identification of $l=\rho_*$. Parameter setting: $L = 1$, $\Omega = 1$, $G_N=1$. For simplicity and higher numerical precision, we have used the first-order derivative $S'(l)$, i.e. the LHS of \eqref{mainformula} instead of the second-order derivative $S''(l)$ which corresponds directly to the CMI.
    }
    \label{fig_EE_HV(2)}
\end{figure}

Then we compare the entanglement entropy data (CMI data) computed as the LHS of \eqref{mainformula} with the bulk geometry data at the radial position $\rho_*$ computed as the RHS of \eqref{mainformula_new}, which are displayed in figure \ref{fig_EE_HV(2)}. The match of the two curves indicates that the bulk geometry at $\rho_*$ determines the boundary CMI behavior at real-space scale $l$. 

\subsection{A special case: bubble shadow in glued geometry}

In this subsection, we discuss a special case, which is the glued geometry of two or more geometries glued at certain radial positions. We take the simple case with two geometries glued at $z=z_0$ as an example. This type of geometry has been studied in the brane world scenario and in holographic interface conformal field theories \cite{Bachas:2020yxv,Simidzija:2020ukv,Liu:2024oxg,Liu:2025gle}. A matter brane exists at $z= z_0$ to make the geometry consistent with Einstein equations. 

In certain glued geometries, there exist a bubble shadow as pointed out in \cite{Burda:2018rpb}, which means that the position of the bulk turning point $z_*$ as a function of the boundary interval $l$ is not a continuous function and has a jump at $z=z_0$ to $z=z_1$ at a critical $l=l_c$. The bulk region which has no turning points between $z=z_0$ and $z=z_1$ is called the bubble shadow. {Therefore, even when an entanglement shadow is not present, there exists a bulk radial range where $z_*$ cannot take these values.} In this case, the coordinate transformation to the new $\rho$-coordinate system is not well defined and the $\rho$-coordinate would become discontinuous. Examples are the BTZ black hole glued with another BTZ black hole in \cite{Burda:2018rpb} and the two AdS$_3$ geometries with different AdS radius glued at an IR radial scale in \cite{Ju:2024xcn}. For boundary CMI at lengths smaller than $l_c$, we can still transform the geometry outside $z=z_0$ to the specific coordinate system where the CMI at length $l$ is {completely} determined by the geometry at the corresponding $z_*(l)$. For the case of \cite{Ju:2024xcn}, the outer geometry is already in this coordinate system with no need to perform a coordinate transformation.

This indicates that the bulk geometry between $z=z_0$ and $z=z_1$ seems to be redundant in determining the dual CMI at certain length scales. In the bulk metric CMI reconstruction proposal in the next section, the bubble shadow geometry indeed poses a problem in the CMI reconstruction. It requires further investigation to elucidate the roles of this part of geometry in the boundary entanglement structure at real-space scales and in the bulk metric reconstruction.

\subsection{Other possible choices of CMI for different subsystems}\label{2.4}

In addition to the specific coordinate system introduced above, other gauge choices of $z_*(l)$ also work, {making no physical difference} when being picked as the criterion for the gauge choice. In this section, we generalize the physics in previous sections in a different direction: we discuss other choices of CMI configurations rather than other choices of $z_*(l)$. More precisely, we consider the boundary CMI for two infinitesimal subsystems with different lengths in contrast with the CMI with the same length in previous sections. This would also result in a correspondence between the bulk geometry and the boundary real-space scale, which involves different metric components in the correspondence. We could follow similar calculations as in the previous subsections to explicitly demonstrate this. However, to avoid unnecessary complexity, we {herein} only provide a sketch of the main picture of this generalization.

\begin{figure}[H]
    \centering
    \includegraphics[width=10.5cm]{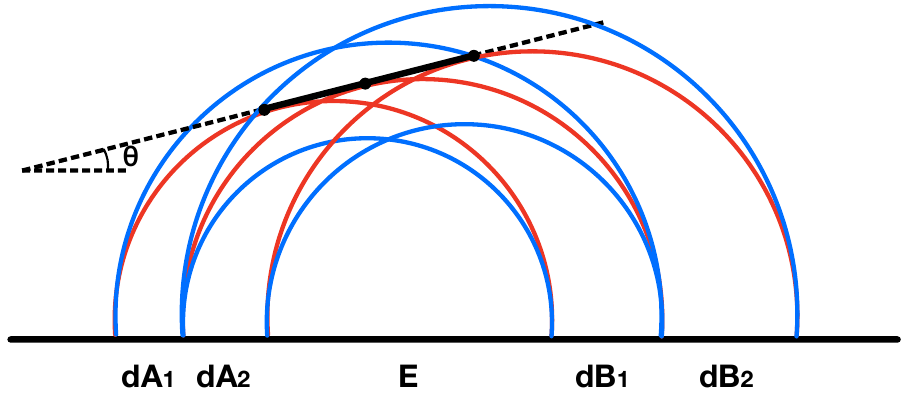}
    \caption{Illustration for the bulk geometry/boundary CMI relation when infinitesimal subregions $A$ ($dA_1$ or $dA_2$) and $B$ ($dB_1$ or $dB_2$) are not equal in size.{ The CMI between $dA_{1(2)}$ and $dB_{1(2)}$ is calculated by the sum of the lengths of left (right) two red curves minus the lengths of the left (right) two blue curves, respectively. Note that $dA_{1(2)}$ and $dB_{1(2)}$ are infinitesimal subregions and all RT surfaces in the figure are infinitely close.} 
    }
    \label{fig_differential_entropy(1)}
\end{figure}

As shown in figure \ref{fig_differential_entropy(1)}, $A$ and $B$ are boundary subregions with $E$ being the interval in between with length $l$. The length of $l_A$ and the length of $l_B$ are not the same anymore, different from the previous sections, {although they are still of infinitesimal lengths}. We pick a point in the bulk at $z=z_{*\theta}$ and introduce an angle $\theta${, where $\theta$ is the coordinate angle of the line tangential to all the red RT surfaces and $z_{*\theta}$ is the $z$ coordinate of the corresponding tangential point. } From the figure, we see that as long as the ratio of $l_B/l_A$ as a function of $\theta$ and $z_{*\theta}$ is chosen accordingly, the CMI $I(A,B|E)$ will be {completely} determined by the first derivative of the metric components  $\sqrt{h(z_\theta)^2 \cos{\theta}^2+f(z_\theta)^2 \sin{\theta}^2}$ in the direction normal to the tangential line. A simple reasoning is that the CMI describes the variation of the differential entropy that corresponds to $S_{AE}-S_{E}$ and $S_{AEB}-S_{EB}$, so its integration gives the differential entropy that corresponds to this metric component. Stated equivalently, an integration of $I(A,B|E)$ in the normal direction, i.e., $dS/dl$, could be totally determined by this combination of metric components. This can be seen from the summation and subtraction of the four geodesic in the definition of $I(A,B|E)$. This can also be shown from that property that the double integration of CMI gives the length of a bulk curve \cite{Czech:2014tva}, which is {the key point of} the differential entropy reconstruction method which we will review in section \ref{3.1}.

{The same as in previous configurations}, which correspond to the $\theta=0$ case {herein}, the correspondence appears to be local. However, the relation between $z_{*\theta}$ and $l$ varies in different coordinate systems, {reflecting} its {nonlocality}. Therefore, {in a similar manner} we can also choose a gauge by fixing a $z_{*\theta}(l)$ function for a given $\theta$, e.g. {$z_*=\frac{l}2 \cos{\theta}$}. Note that there is a subtle difference for nonzero $\theta$ from the $\theta=0$ case: for a fixed $\theta$ and a given gauge $z_{*\theta}(l)$, to obtain the corresponding boundary CMI $I(A,B|E)$ and its relation with the bulk metric components, {it requires us} to know the ratio $l_B/l_A$, which is one in the case of $\theta=0$. This ratio {is computable} when we know the geometry for a given $\theta$ and $z_{*\theta}$ by plotting the two tangential geodesics at two adjacent points at $z_{*\theta}$, and then investigating the ratio of the distance between the points where those two tangential geodesics intersect with the boundary. This does not change the result that the bulk radial geometry/boundary CMI correspondence still appears to be local for nonzero $\theta$. However, this poses a challenge when we {attempt} to reverse this process to reconstruct the bulk metric from boundary CMI data since more data (the ratio) would be required. 

\section{A new method of bulk metric reconstruction: the algorithm}\label{3}

\noindent Inspired from the calculation in the previous section, we propose a new method for bulk metric reconstruction, namely the CMI reconstruction, which is the inverse operation of section \ref{2}. In this section, we will first review the Bilson's reconstruction procedure \cite{Bilson:2010ff} and the differential entropy reconstruction \cite{Czech:2014ppa,Czech:2014tva,Burda:2018rpb}. Then we introduce the CMI reconstruction, {with typical examples provided in section \ref{4}}. We will show that this new CMI reconstruction method is relatively easier and more computationally efficient, {bridging between} the differential entropy reconstruction and the Bilson's reconstruction method.

\subsection{Review of Bilson's reconstruction and the differential entropy reconstruction}\label{3.1}

Bilson's method, especially suitable in the gauge choice of the metric function $h(z)=1$\footnote{Note that in the bulk metric reconstruction from boundary entanglement data, the time component $g_{tt}$ of the metric does not have any effect, so in this review of bulk metric reconstructions and our CMI reconstruction, we will not reconstruct the time component of the metric. Then a gauge of $h(z)=1$ could in general be picked. }, provides a shortcut to reconstruct the other metric function $f(z)$ using the scaling law of the entanglement entropy $S(l)$.

Going back to the $z$-coordinate system, recall the relationship of $S(l)$ and $h(z_*)$ in \eqref{mainformula}:
\begin{equation}
    \frac{dS}{dl}=\frac{dz_*}{dl}\frac{dS}{dz_*}=\frac{L^2\Omega}{4G_N}\frac{h(z_*)}{z_*^2},
\end{equation}
where the LHS is now a known function of the boundary length $l$. Therefore, using this identity, one can determine the relation of $l$ and the bulk radial position $z_*$ with $h(z)=1$. 

At the same time, one can prove that solving \eqref{lzstar} would yield a solution
\begin{equation}
    \sqrt{\frac{1}{f(z)h(z)^3}} = \frac{2}{\pi z^2}\frac{d}{dz}\int_0^z
    \frac{l(z_*)}{\sqrt{\frac{z^4}{h(z)^2}-\frac{z_*^4}{h(z_*)^2}}}
    \frac{2z_*^3h(z_*)-z_*^4h'(z_*)}{2h(z_*)^3}\;dz_*.
\end{equation}
Insert the relation $l(z_*)$ obtained from (3.1), and one can simply reconstruct $f(z)$ with the help of this solution with $h(z)=1$. Equivalently, one can also use the relation $S(z_*)$ in \eqref{HEE_AdS4} to solve $f(z)$ as a substitute for \eqref{lzstar}.

Next, we review the differential entropy reconstruction proposed in \cite{Czech:2014ppa}. Following \cite{Czech:2014ppa}, we use $\tilde{\theta}$ and $R$ to represent the bulk coordinate. As an example, the metric of pure AdS$_3$ in static coordinates is: 

\begin{equation}\label{metric_pure_AdS3}
   ds^2=-\left(1+\frac{R^2}{L^2}\right)dt^2+\left(1+\frac{R^2}{L^2}\right)^{-1}dR^2+R^2d\tilde\theta^2.
\end{equation}

\begin{figure}[H]
    \centering
    \includegraphics[width=\textwidth]{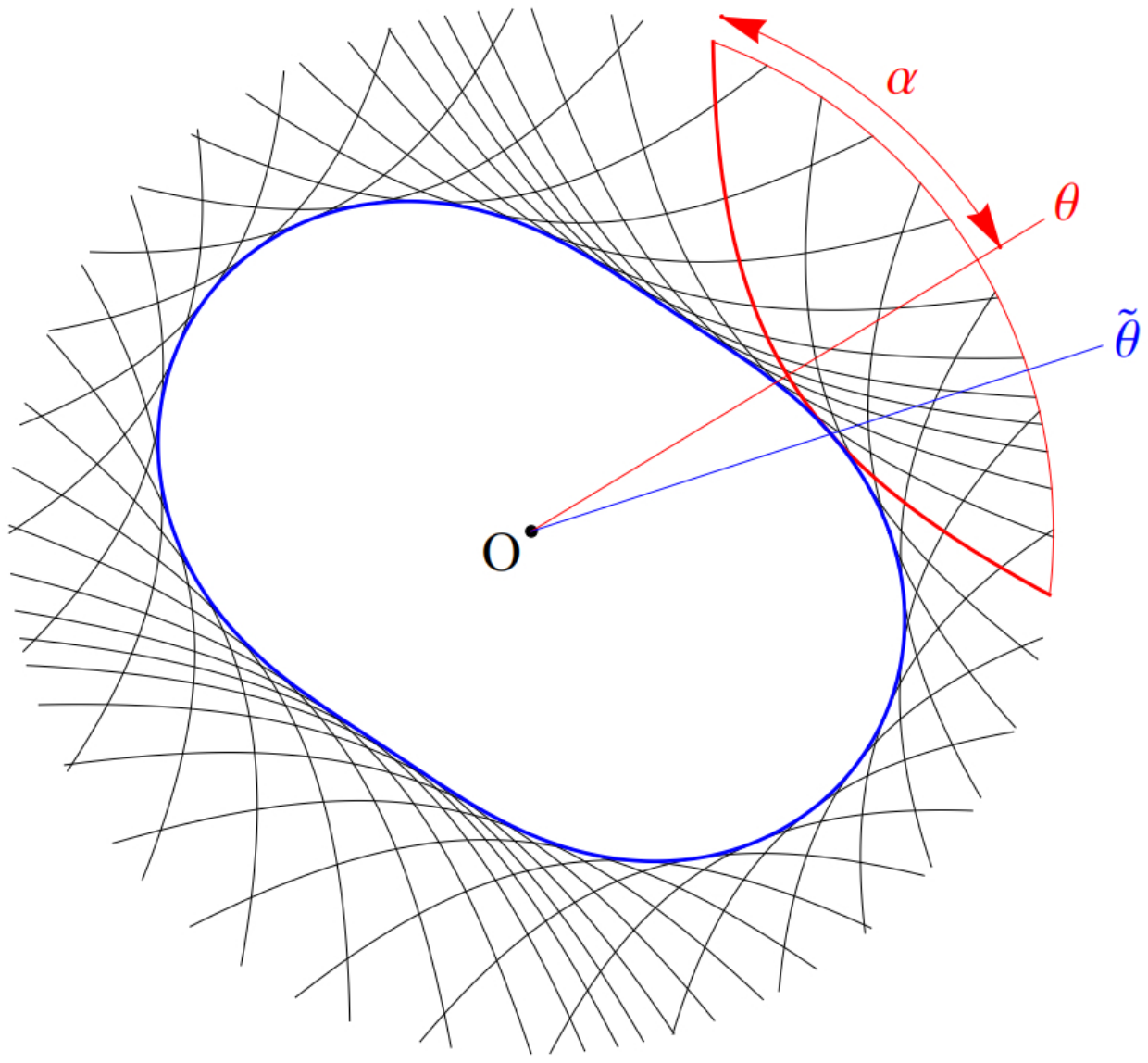}
    \caption{The blue curve is {an} arbitrary closed smooth curve in the bulk $R=R(\tilde{\theta})$. Black spacelike geodesics are tangent to the given blue curve. The geodesics are centered at $\theta(\tilde{\theta})$ and have width $2\alpha(\theta)$. {This figure is taken from \cite{Czech:2014ppa}.}
    }
    \label{fig_differential_entropy(2)}
\end{figure}

As shown in figure \ref{fig_differential_entropy(2)},
Consider a closed, smooth bulk curve $R=R(\tilde{\theta})$ on the $T=0$ slice. For any point on the curve, we {can} find a geodesic tangent to the curve at that point. We use $\theta(\tilde{\theta})$ to represent the angular coordinate of the center of that geodesic, with $\tilde{\theta}$ denoting the tangency point on the curve $R(\tilde{\theta})$. Note that this $\theta$ is different from the one defined in section \ref{2.4}. 

We denote the endpoints of the geodesic with $\theta({\tilde\theta})\pm\alpha(\theta)$. That is to say, the width of the geodesic is $2\alpha(\tilde\theta)$. We find that the curve $R(\tilde{\theta})$ is associated with a boundary function $\alpha(\theta)$. Then the circumference of the curve $R(\tilde{\theta})$ can be computed using differential entropy denoted by $E[\alpha(\theta)]$\cite{Balasubramanian:2013lsa},

\begin{equation}
    E[\alpha(\theta)]=\frac12\int_0^{2\pi}d\theta\frac{dS(\alpha)}{d\alpha}|_{\alpha=\alpha(\theta)}=\frac{circumference}{4G_N}.
\end{equation}

Thus, we {can} reconstruct the circumference of the bulk curve $R(\tilde{\theta})$ from the boundary function 
$\alpha(\theta)$ using differential entropy, which is a combination of derivatives of boundary entanglement entropies.

   But to reconstruct the metric of the bulk, that is not enough. Therefore, we now consider how to reconsruct points in the bulk.
In a space with negative curvature, we have the Gauss-Bonnet theorem:
\begin{equation}\label{GB}
\oint_Cd\tau\sqrt{h}K = 2\pi-\int_A RdA\geq2\pi,
\end{equation}
where $C$ is an {arbitrary} closed curve, $A$ is the area enclosed by the curve $C$, and $R$ is the Ricci scalar.
If a closed curve in this space {shrinks} to a point, the integral in \eqref{GB} approaches the extremal value of $2\pi$.
Thus we can pick out any point in this space as the extrema of the extrinsic curvature.
Asymptotic AdS$_3$ spacetime is also a {negatively} curved spacetime. Therefore, utilizing the following identity
\begin{equation}
d\tau\sqrt{h}K=d\theta\sqrt{-\frac{d^2S}{d\alpha^2}(1-\alpha'(\theta)^2)},
\end{equation}
for any point $A$ in an asymptotically AdS$_3$ spacetime, we can use the Euler-Lagrange equation to compute the boundary function $\alpha_A(\theta)$ associated with the point $A$ 
\begin{equation}
(2\alpha_A'')\frac{d^2S}{d\alpha^2}|_{\alpha_A}+(1-\alpha_A'^2)
\frac{d^3S}{d\alpha^3}|_{\alpha_A}=0.
\end{equation} This means that we {can} reconstruct the bulk point $A$ from the boundary function $\alpha_A(\theta)$.

Then we move on to the reconstruction of distance in differential entropy reconstruction. If we have two bulk points $A$ and $B$, their convex cover is the ``closed curve” that runs from $A$ to $B$ along a geodesic and then returns from $B$ to $A$. The circumference of the cover, which is twice the distance between $A$ and $B$, is then given by $E[\gamma(\theta)]$, where $\gamma(\theta)$ is $min\{\alpha_A(\theta),\alpha_B(\theta)\}$ for all $\theta$. Now, we can define the geodesic distance between $A$ and $B$ with associated boundary functions $\alpha_A(\theta)$ and $\alpha_B(\theta)$
\begin{equation}\label{geodesic_distance}
d(A,B)=\frac12E[\gamma(\theta)].
\end{equation}

We verify this result with the pure AdS$_3$ example. In pure AdS$_3$, the associated boundary function $\alpha_A(\theta)$ of a point $A$ in the bulk is
\begin{equation}
\alpha_A(\theta)=\cos^{-1}\frac{R\cos(\theta-\tilde\theta)}{\sqrt{L^2+R^2}}.
\end{equation}
{Therefore,} we can compute the distance between the point $A$ at $R_A=R$ and $\tilde{\theta}_A=0$ on the $T = 0$ slice of AdS$_3$ and the origin $O$
\begin{equation}\label{distance_to_origin}
d(O,A)=\frac{L}{8G_N}\int_{-\pi/2}^{\pi/2}d\theta
\frac{R\cos\theta}{\sqrt{L^2+R^2\sin^2\theta}}
=\frac1{4G_N}\int_0^Rdr(1+\frac{r^2}{L^2})^{-1/2}.
\end{equation}
It is obvious that the result in \eqref{distance_to_origin} is consistent with \eqref{metric_pure_AdS3}.

\subsection{CMI bulk metric reconstruction}\label{3.2}

In section \ref{2}, we have shown that in a specific coordinate system, the boundary behavior of the CMI of two small subsystems at a distance $l$ would be totally determined by the bulk geometry at $\rho=l$ under a special gauge choice, from general calculations and several explicit examples. This inspires us to perform an inverse calculation, i.e. to reconstruct the spatial components of the bulk metric as the output by having the behavior of the CMI as the input. This {is practicable} in a specific coordinate system. With the result of the {reconstructed} bulk metric in the special coordinate system, we can further transform it to any other coordinate system of interest.

A summary of the strategy goes as follows.

\begin{itemize}
    \item Step 1: reconstruct metric function $H(\rho)$.
\end{itemize}

Given the scale dependence of the entanglement entropy $S(l)$, or equivalently $S'(l)$ or $S''(l)$ which is proportional to the CMI, a reformulation of \eqref{mainformula} in the $\rho$-coordinate system yields
\begin{equation}\label{reconstruction_H}
    \frac{dS}{dl}=\frac{dS}{d\rho_*}=\frac{L^2\Omega}{4G_N}\frac{H(\rho_*)}{\rho_*^2},
\end{equation}
from which one can simply solve $H(\rho_*)$ with the identification of $l=\rho_*$ in the particular gauge choice. Going through all values of $\rho_*$ or $l$ equivalently, one can obtain the function $H(\rho)$. 

\begin{itemize}
    \item Step 2: reconstruct metric function $F(\rho)$.
\end{itemize}

Once $H(\rho)$ is obtained, we use \eqref{lzstar} to retrieve $F(\rho)$. First, we reformulate this equation in the $\rho$-coordinate system as
\begin{equation}
    l=\rho_*=2\int_0^{\rho_*}d\rho \frac{1}{\sqrt{\frac{H(\rho)^2 \rho_*^4}{H(\rho_*)^2 \rho^4}-1}}\frac{1}{\sqrt{H(\rho)F(\rho)}}.
\end{equation}
One can prove that the solution is \footnote{\cite{Ahn:2024jkk} derived the solutions in  $z$-coordinate system.}
\begin{equation}\label{reconstruction_F}
    \sqrt{\frac{1}{F(\rho)H(\rho)^3}} = \frac{2}{\pi\rho^2}\frac{d}{d\rho}\int_0^\rho
    \frac{\rho_*}{\sqrt{\frac{\rho^4}{H(\rho)^2}-\frac{\rho_*^4}{H(\rho_*)^2}}}
    \frac{2\rho_*^3H(\rho_*)-\rho_*^4H'(\rho_*)}{2H(\rho_*)^3}\;d\rho_*,
\end{equation}
from which $F(\rho)$ can be derived. 

{With the bulk metric reconstructed} from boundary CMI data ($S''(l)$) or equivalently $S'(l)$ following these two steps, we can further perform a coordinate transformation to any other coordinate system {of interest}. Using this coordinate transformation, we also examine the correctness of the CMI reconstruction method {in} the following way. Assume in the $z$-coordinate system in which we have obtained the CMI data, the corresponding metric components are $h(z)$ and $f(z)$, {respectively}. We use the coordinate transformation procedure performed in section \ref{2} to obtain the metric components in $\rho$- coordinate system with the gauge $\rho_*=l$, which become $H_{target}(\rho)=\frac{\rho^2}{z(\rho)^2}h(z(\rho))$ and $F_{target}(\rho)=\frac{z(\rho)^2}{z'(\rho)^2\rho^2}f(z(\rho))$, respectively. We verify the correctness of the CMI reconstruction method by showing $H(\rho)=H_{target}(\rho)$ and $F(\rho)=F_{target}(\rho)$. In section \ref{4}, we will perform the two steps of CMI bulk metric reconstruction in several examples and use the coordinate transformation mentioned above to {verify} the correctness of the reconstruction.

\subsection{Connections between CMI reconstruction method and previous methods}

The CMI reconstruction can be viewed as a specific gauge choice in the Bilson's reconstruction method where the latter is a general framework for all possible coordinate systems. {Note that the CMI prescription is based on the RT prescription and consequently suffers from the same limitation that the geometry in the entanglement shadow cannot be reconstructed.}

From (\ref{mainformula}), we see that the integration function of the CMI ($S'(l)$) depends on the metric locally {instead of} being an integration function {depending} on the metric in the UV region. Therefore, the differential entropy reconstruction could be viewed as the integration of CMI {while} the integration depends on the geometry locally. Consequently, the differential entropy reconstruction is an integration version of the CMI reconstruction and is gauge invariant, while the latter is a local version of the differential entropy reconstruction and depends on gauge choices. The dependence comes from $d z_*/d l$ when we take the derivative of $S'(l)$ with $l$. Each configuration in the integration of the differential entropy reconstruction corresponds to one possibility of the CMI reconstruction with different $\theta$ introduced in section \ref{2.4}. Therefore, differential entropy reconstruction is an integration version of the CMI reconstruction and the latter could be viewed as the former decomposed in each gauge. Therefore, the CMI reconstruction can be regarded as a bridge connecting the general Bilson's reconstruction algorithm and the differential entropy method.

\section{Examples of the CMI bulk metric reconstruction}\label{4}

In this section, we provide explicit examples of metric reconstruction using our CMI prescription, which are examples from the same systems as in section \ref{2}. We will then confirm the correctness of the procedure using the computation in section \ref{2}.

\subsection{AdS black hole}

Our reconstruction computation begins with the case of AdS black hole, aiming to retrieve the data of $H(\rho)$ and $F(\rho)$ in the $\rho$-coordinate system with gauge $\rho_*=l$ from the CMI data of $S''(l)$ or equivalently $S'(l)$ which we already have in advance. 

Following our reconstruction strategy in section \ref{3.2}, first we reconstruct $H(\rho)$ according to \eqref{reconstruction_H}, with the entanglement entropy function $S(l)$ (or  equivalently $S'(l)$ or the CMI ($-S''(l)$)) obtained in section \ref{2.2.3} used as the input. The reconstructed $H(\rho)$ is obtained numerically using the CMI reconstruction procedure. Then we reconstruct $F(\rho)$ according to \eqref{reconstruction_F}, with the reconstructed $H(\rho)$ used as the input. Then with these two reconstructed functions $H(\rho)$ and $F(\rho)$, we further {verify the correctness of reconstruction} by comparing them with the metric components $H_{target}(\rho)$ and $F_{target}(\rho)$ which are directly obtained from the coordinate transformation from the original $z$-coordinate system for the AdS black hole. As long as these two sets of functions match with each other, the {validity} of the CMI reconstruction can be confirmed. 

\begin{figure}[!htbp]
    \centering
     \includegraphics[width=\textwidth]{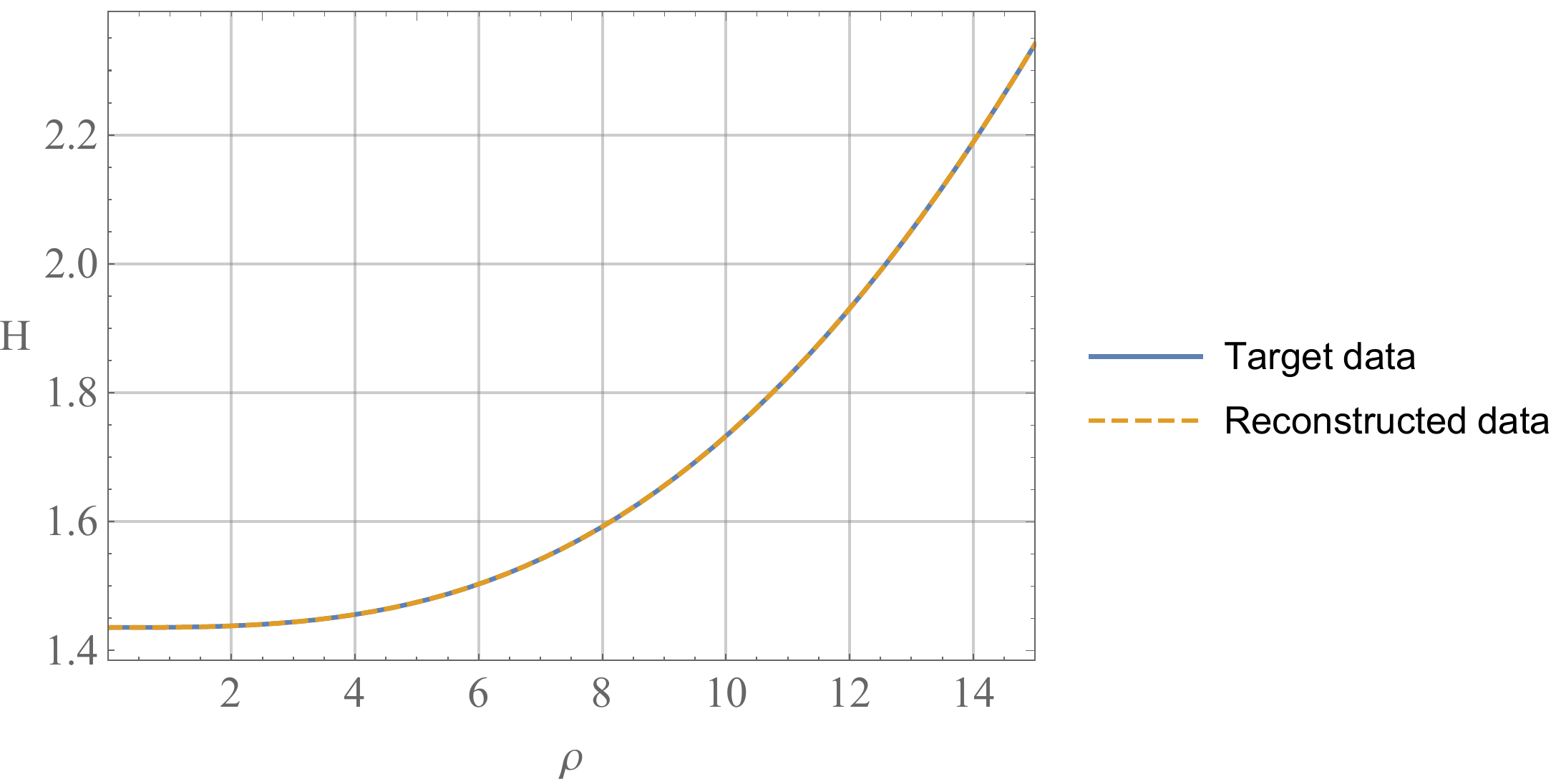}
    \caption{The reconstructed $H(\rho)$ (dashed yellow curve) and the target function $H_{target}(\rho)$ (solid blue curve) in AdS black hole geometry. Parameter setting: $L = 1$, $\Omega = 1$, $G_N=1$.}
    \label{fig_H_reconstruct_AdSBH}
\end{figure}

\begin{figure}[!htbp]
    \centering
    \includegraphics[width=\textwidth]{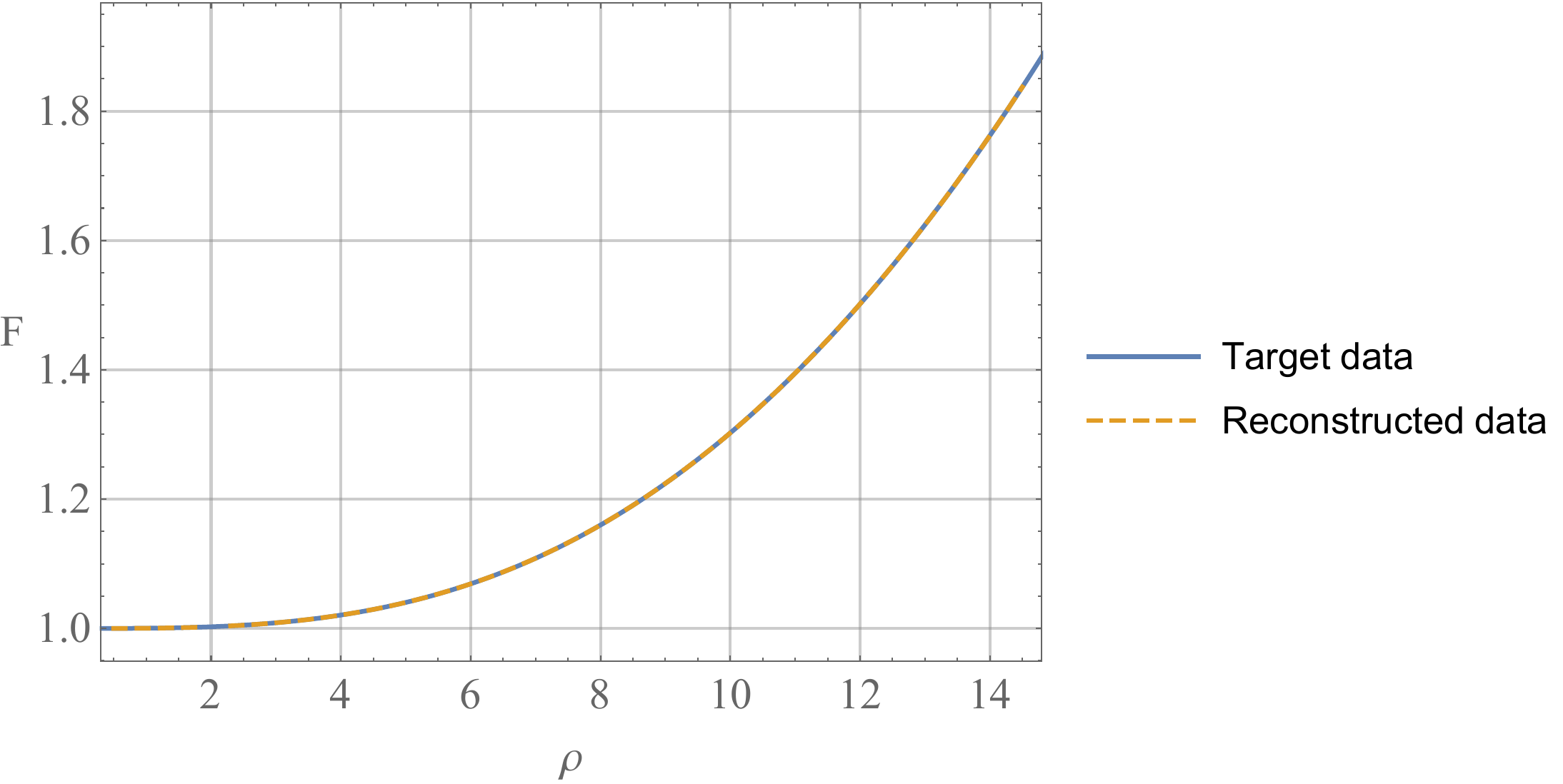}
    \caption{The reconstructed $F(\rho)$ (dashed yellow curve) and the target function $F_{target}(\rho)$ (solid blue curve) in AdS black hole geometry.}
    \label{fig_F_reconstruct_AdSBH}
\end{figure}

Figure \ref{fig_H_reconstruct_AdSBH} shows the reconstructed function $H(\rho)$
and the target function $H_{target}(\rho)$ directly obtained from the coordinate transformation. The reconstructed $F(\rho)$ and the target function $F_{target}(\rho)$ are displayed in figure \ref{fig_F_reconstruct_AdSBH}. It is manifest from figure \ref{fig_H_reconstruct_AdSBH} and \ref{fig_F_reconstruct_AdSBH} that our reconstructed data match perfectly with the target data.


\subsection{BTZ black hole}

Using the CMI bulk reconstruction procedure,  $H(\rho)$ can be obtained from the relationship (\ref{mainformula}) between the derivative of the entanglement entropy and the metric
\begin{equation}
\frac{dS}{dl} = \frac{L}{4G}\sqrt{\frac{H(\rho_*)}{\rho_*^{2}}}.
\end{equation}
Then $F(\rho)$ can be inversely solved from (\ref{lzstar}), and it is the same as the $F(\rho)$ obtained through coordinate transformation
\begin{equation}
    \sqrt{\frac{1}{F(\rho)}} = \frac{H(\rho)}{2\pi\rho}\frac{d}{d\rho}\int_0^\rho\frac{l(\rho_*)}{\sqrt{\frac{\rho^2}{H(\rho)}-\frac{\rho_*^2}{H(\rho_*)}}}\frac{2\rho_*H(\rho_*)-\rho_*^2H'(\rho_*)}{H(\rho_*)^2}d\rho_*.
\end{equation}
From analytic calculations, we can check that these two functions match with the target functions that we have already obtained in section \ref{2} as follows

\begin{equation}
    H_{target}(\rho) = \frac{\rho^2}{z^2}h(z)  = \left(\frac{\sqrt{M}\rho}{\tanh{\frac12\sqrt{M}\rho}}\right)^2,
\end{equation}
and 
\begin{equation}
    F_{target}(\rho) = \frac{z^2f(z)}{\rho^2z'(\rho)^2}  = \left(\frac{2\sinh(\frac12\sqrt{M}\rho)}{\sqrt{M}\rho}\right)^2.
\end{equation}

\subsection{A flow geometry in Einstein-Maxwell-dilaton theory}

\begin{figure}[!htbp]
    \centering
    \includegraphics[width=\textwidth]{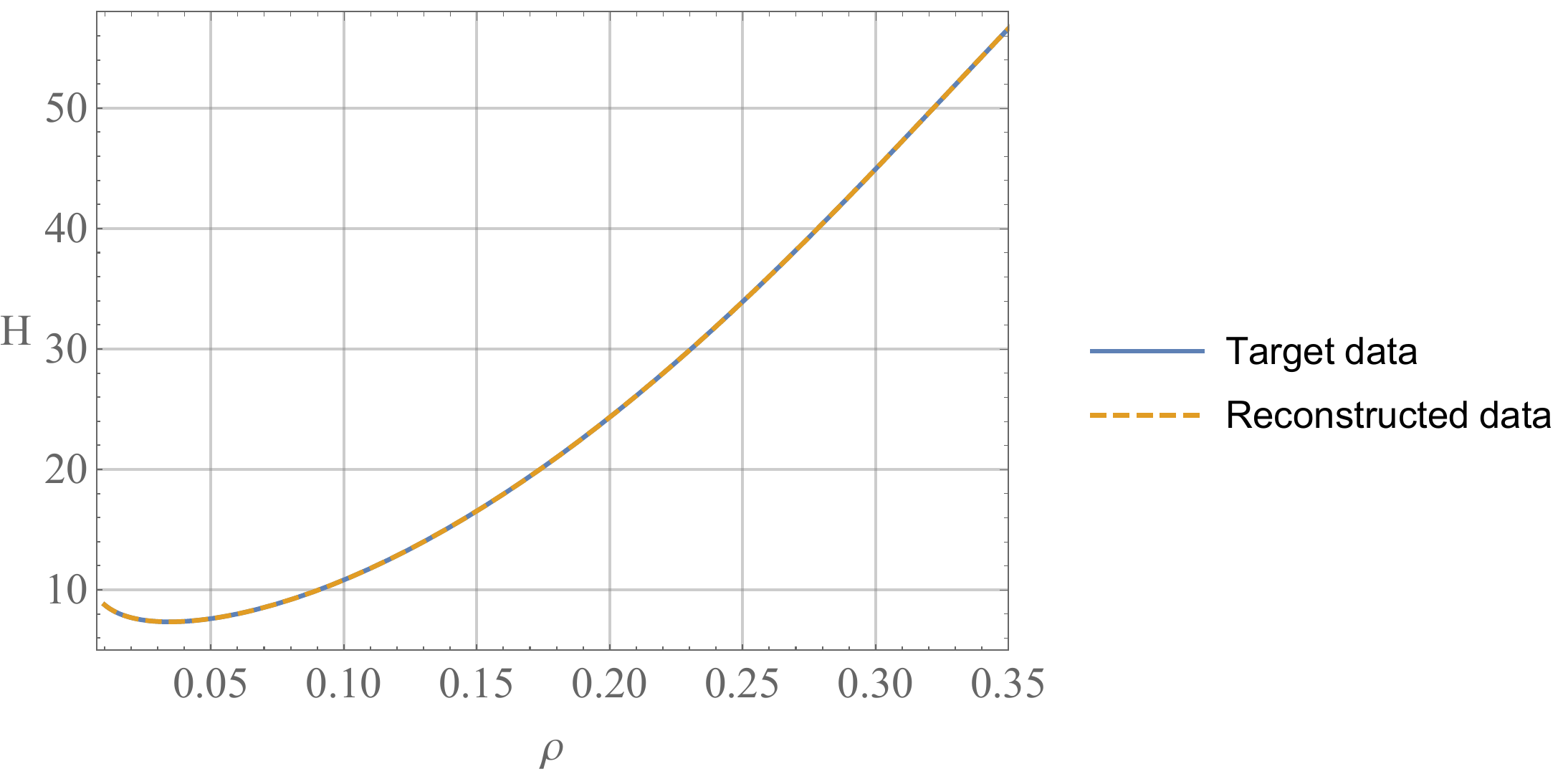}
    \caption{The reconstructed $H(\rho)$ (dashed yellow curve) and the target function $H_{target}(\rho)$ (solid blue curve) in the flow geometry. Parameter setting: $L = 1$, $\Omega = 1$, $G_N=1$. In this geometry, $\rho = 0.35$ corresponds to $z \approx 100$ in the old coordinate system.}
    \label{fig_H_reconstruct_HV}
\end{figure}

\begin{figure}[!htbp]
    \centering
    \includegraphics[width=\textwidth]{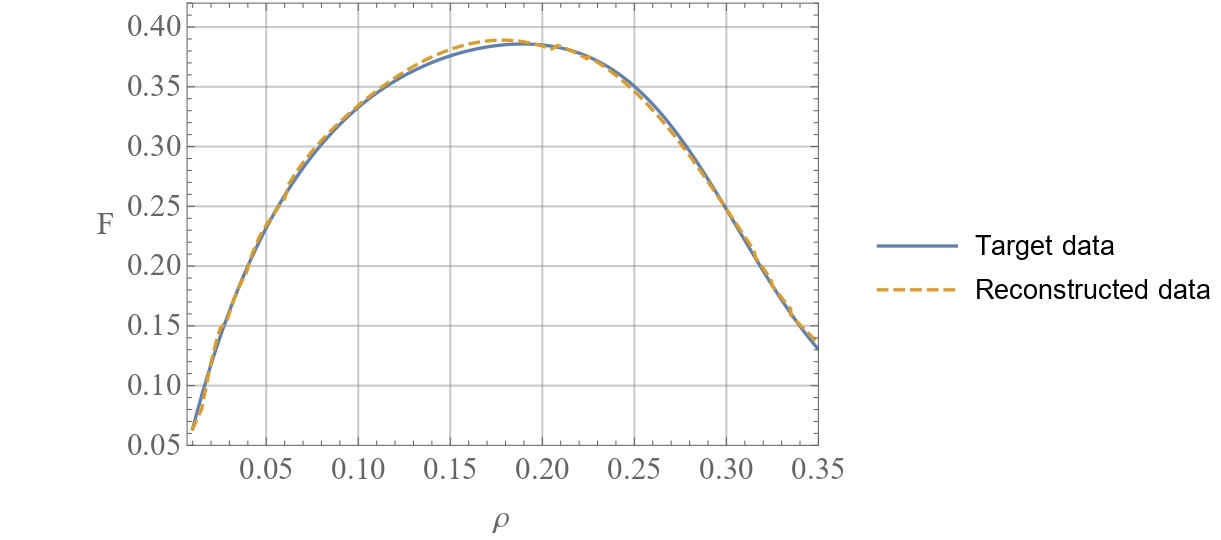}
    \caption{The reconstructed $F(\rho)$ (dashed yellow curve) and the target function $F_{target}(\rho)$ (solid blue curve) in the flow geometry. In this geometry, $\rho = 0.35$ corresponds to $z \approx 100$ in the old coordinate system.}
    \label{fig_F_reconstruct_HV}
\end{figure}

In this final example, we turn to the case of the flow geometry studied in section \ref{2.2.4}. Following our reconstruction strategy in section \ref{3.2}, first we reconstruct $H(\rho)$ according to \eqref{reconstruction_H}, with the entanglement entropy $S(l)$ or equivalently the CMI ($-S''(l)$) obtained in section \ref{2.2.4} used as the input boundary data. The reconstructed $H(\rho)$ and the target function $H_{target}(\rho)$ directly obtained from the coordinate transformation are displayed in figure \ref{fig_H_reconstruct_HV}. Next, we reconstruct $F(\rho)$ according to \eqref{reconstruction_F}, with the reconstructed $H(\rho)$ used as the input. The reconstructed $F(\rho)$ and the target function $F_{target}(\rho)$ are displayed in figure \ref{fig_F_reconstruct_HV}.

These two sets of functions again match with each other. Note that the computational error in the reconstructed $H(\rho)$ data is negligible while the error in the reconstructed $F(\rho)$ data is not infinitesimal. Such kind of error exists inevitably in the case of generally numerical background spacetime data due to machine precision, e.g. the numerical solution of $a(r)$, $b(r)$, and $\phi(r)$ in our case, while {it shall neither alter the physical picture or undermine the feasibility of the CMI reconstruction method}. 


\section{Conclusion and discussion}\label{5}

In this work we have examined deeper into the exact relation between the boundary entanglement structure of the CMI for two infinitesimal subsystems at a distance $l$ and the bulk geometry at a corresponding radial position $z_*$ which determines this behavior. There appears to be a local one-to-one correspondence between $z_*$ and $l$, however, the relation $z_*(l)$, i.e. the geometry at which the radial position determines the boundary CMI at real-space scale $l$, depends on gauge choices of coordinate systems. This reflects the nonlocal property of gravity as the boundary CMI behavior should be determined by the geometry in the whole UV region outside $z_*$. Therefore, this bulk radial geometry/boundary CMI correspondence only appears to be a local one-to-one correspondence  and the relation $z_*(l)$ varies in different gauge choices. 

We propose to use the function of $z_*(l)$ as the gauge choice, which both fixes the gauge and establishes a priori the relationship between the boundary real-space scale $l$ and the bulk radial scale $z_*$. We present the exact bulk radial geometry at $z_*$/boundary CMI behavior at real-space scale $l$ correspondence in the simplest gauge $z_*=l$ and provide several examples showing this explicitly, 
including a geometry flowing from AdS$_2$ in the near-horizon region to AdS$_4$ at the boundary. We emphasize again that the bulk radial geometry/boundary CMI behavior correspondence only appears to be local and is {in essence} nonlocal as it depends on gauge choices.

Inspired by this correspondence, we also propose a new bulk metric reconstruction method, namely the CMI reconstruction. We reconstruct the bulk metric from boundary CMI data in a give gauge choice of $z_*(l)$ and then we could perform a coordinate transformation to any other coordinate system of interest. The CMI reconstruction bridges between the Bilson's general algorithm and the differential entropy prescription. One open question is that in this work we have focused on geometries with translational symmetry and it would be interesting to see if this could be generalized to geometries without translational symmetry.

In this work, to justify the bulk geometry/boundary CMI correspondence and to formulate the bulk metric reconstruction procedure, we have focused on effectively one-dimensional boundary entangling regions, i.e. strips. It would be interesting to study the higher-dimensional generalizations and examine the validity of the correspondence in these cases, e.g. in spherical regions. {Furthermore, it would be interesting to generalize the present CMI prescription to anisotropic bulk metrics utilizing higher dimensional boundary entanglement data, for which more generic shapes of boundary regions other than strips need to be considered.}

{From another perspective, we have not reconstructed the $g_{tt}$ component. In principle, the $g_{tt}$ component can be reconstructed via the procedure detailed below. Consider a boosted Cauchy slice and perform the CMI reconstruction in this boosted Cauchy slice using the prescription of this work. On any boosted Cauchy slice, the metric components tangent to the slice can be readily recovered; the normal component, however, is inaccessible from a single boost. To obtain the complete metric including the $g_{tt}$ component, one should combine the data of the two spacelike regions: one on the boosted slice and the other on the original unboosted slice. The price for this reconstruction is that one needs the two independent data sets plus precise knowledge of the boost transformation.}

Using the CMI method, it would also be an important problem to investigate potential extensions of current metric reconstruction results to broader scenarios where the spacetime breaks the translational symmetry, e.g. the ansatz in \cite{Jokela:2025ime}, which we will report in a future work. Possible input data could be a combination of CMI (or entanglement entropy) and entanglement wedge cross section considering that the metric functions ($f$ and $h$) would exhibit a simultaneous dependence on the radial coordinate $r$ and spatial coordinate $x$.

\section*{Acknowledgement}

We thank Bartlomiej Czech, Keun-Young Kim, Li Li, Bo-Hao Liu and Yan Liu for helpful discussions. This work is supported by National Natural Science Foundation of China (Grant No. 12408058, No. 12035016, and No. 12247103).


\bibliographystyle{elsarticle-num}
\bibliography{hyperref}

\begin{thebibliography}{10}
\expandafter\ifx\csname url\endcsname\relax
  \def\url#1{\texttt{#1}}\fi
\expandafter\ifx\csname urlprefix\endcsname\relax\def\urlprefix{URL }\fi
\expandafter\ifx\csname href\endcsname\relax
  \def\href#1#2{#2} \def\path#1{#1}\fi

\bibitem{Maldacena:1997re}
J.~M. Maldacena, {The large N limit of superconformal field theories and supergravity}, Adv.Theor.Math.Phys. 2 (1998) 231--252.
\newblock \href {http://arxiv.org/abs/hep-th/9711200} {\path{arXiv:hep-th/9711200}}, \href {https://doi.org/10.1023/A:1026654312961, 10.1023/A:1026654312961} {\path{doi:10.1023/A:1026654312961, 10.1023/A:1026654312961}}.

\bibitem{Ryu:2006bv}
S.~Ryu, T.~Takayanagi, {Holographic derivation of entanglement entropy from AdS/CFT}, Phys. Rev. Lett. 96 (2006) 181602.
\newblock \href {http://arxiv.org/abs/hep-th/0603001} {\path{arXiv:hep-th/0603001}}, \href {https://doi.org/10.1103/PhysRevLett.96.181602} {\path{doi:10.1103/PhysRevLett.96.181602}}.

\bibitem{Czech:2012bh}
B.~Czech, J.~L. Karczmarek, F.~Nogueira, M.~Van~Raamsdonk, {The gravity dual of a density matrix}, Class. Quant. Grav. 29 (2012) 155009.
\newblock \href {http://arxiv.org/abs/1204.1330} {\path{arXiv:1204.1330}}, \href {https://doi.org/10.1088/0264-9381/29/15/155009} {\path{doi:10.1088/0264-9381/29/15/155009}}.

\bibitem{Wall:2012uf}
A.~C. Wall, {Maximin surfaces, and the strong subadditivity of the covariant holographic entanglement entropy}, Class. Quant. Grav. 31~(22) (2014) 225007.
\newblock \href {http://arxiv.org/abs/1211.3494} {\path{arXiv:1211.3494}}, \href {https://doi.org/10.1088/0264-9381/31/22/225007} {\path{doi:10.1088/0264-9381/31/22/225007}}.

\bibitem{Headrick:2014cta}
M.~Headrick, V.~E. Hubeny, A.~Lawrence, M.~Rangamani, {Causality \& holographic entanglement entropy}, JHEP 12 (2014) 162.
\newblock \href {http://arxiv.org/abs/1408.6300} {\path{arXiv:1408.6300}}, \href {https://doi.org/10.1007/JHEP12(2014)162} {\path{doi:10.1007/JHEP12(2014)162}}.

\bibitem{Bousso:2022hlz}
R.~Bousso, G.~Penington, {Entanglement wedges for gravitating regions} (8 2022).
\newblock \href {http://arxiv.org/abs/2208.04993} {\path{arXiv:2208.04993}}.

\bibitem{Espindola:2018ozt}
R.~Esp\'\i{}ndola, A.~Guijosa, J.~F. Pedraza, {Entanglement wedge reconstruction and entanglement of purification}, Eur. Phys. J. C 78~(8) (2018) 646.
\newblock \href {http://arxiv.org/abs/1804.05855} {\path{arXiv:1804.05855}}, \href {https://doi.org/10.1140/epjc/s10052-018-6140-2} {\path{doi:10.1140/epjc/s10052-018-6140-2}}.

\bibitem{Saraswat:2020zzf}
K.~Saraswat, N.~Afshordi, {Extracting Hawking radiation near the horizon of AdS black holes}, JHEP 02 (2021) 077.
\newblock \href {http://arxiv.org/abs/2003.12676} {\path{arXiv:2003.12676}}, \href {https://doi.org/10.1007/JHEP02(2021)077} {\path{doi:10.1007/JHEP02(2021)077}}.

\bibitem{dong2016reconstruction}
X.~Dong, D.~Harlow, A.~C. Wall, Reconstruction of bulk operators within the entanglement wedge in gauge-gravity duality, Physical review letters 117~(2) (2016) 021601.

\bibitem{Leutheusser:2022bgi}
S.~Leutheusser, H.~Liu, {Subalgebra-subregion duality: emergence of space and time in holography} (12 2022).
\newblock \href {http://arxiv.org/abs/2212.13266} {\path{arXiv:2212.13266}}.

\bibitem{Ju:2024xcn}
X.-X. Ju, T.-Z. Lai, B.-H. Liu, W.-B. Pan, Y.-W. Sun, {Entanglement structures from modified IR geometry}, JHEP 07 (2024) 181.
\newblock \href {http://arxiv.org/abs/2404.02737} {\path{arXiv:2404.02737}}, \href {https://doi.org/10.1007/JHEP07(2024)181} {\path{doi:10.1007/JHEP07(2024)181}}.

\bibitem{Ju:2023bjl}
X.-X. Ju, W.-B. Pan, Y.-W. Sun, Y.-T. Wang, {Generalized Rindler wedge and holographic observer concordance} (2 2023).
\newblock \href {http://arxiv.org/abs/2302.03340} {\path{arXiv:2302.03340}}.

\bibitem{Ju:2023dzo}
X.-X. Ju, B.-H. Liu, W.-B. Pan, Y.-W. Sun, Y.-T. Wang, {Squashed entanglement from generalized Rindler wedge} (10 2023).
\newblock \href {http://arxiv.org/abs/2310.09799} {\path{arXiv:2310.09799}}.

\bibitem{Balasubramanian:2013rqa}
V.~Balasubramanian, B.~Czech, B.~D. Chowdhury, J.~de~Boer, {The entropy of a hole in spacetime}, JHEP 10 (2013) 220.
\newblock \href {http://arxiv.org/abs/1305.0856} {\path{arXiv:1305.0856}}, \href {https://doi.org/10.1007/JHEP10(2013)220} {\path{doi:10.1007/JHEP10(2013)220}}.

\bibitem{Ju:2024hba}
X.-X. Ju, W.-B. Pan, Y.-W. Sun, Y.~Zhao, {Holographic multipartite entanglement from the upper bound of $n$-partite information} (11 2024).
\newblock \href {http://arxiv.org/abs/2411.07790} {\path{arXiv:2411.07790}}.

\bibitem{Ju:2024kuc}
X.-X. Ju, W.-B. Pan, Y.-W. Sun, Y.-T. Wang, Y.~Zhao, {More on the upper bound of holographic n-partite information}, JHEP 03 (2025) 184.
\newblock \href {http://arxiv.org/abs/2411.19207} {\path{arXiv:2411.19207}}, \href {https://doi.org/10.1007/JHEP03(2025)184} {\path{doi:10.1007/JHEP03(2025)184}}.

\bibitem{Xu:2023eof}
W.-B. Xu, S.-F. Wu, {Reconstructing black hole exteriors and interiors using entanglement and complexity}, JHEP 07 (2023) 083.
\newblock \href {http://arxiv.org/abs/2305.01330} {\path{arXiv:2305.01330}}, \href {https://doi.org/10.1007/JHEP07(2023)083} {\path{doi:10.1007/JHEP07(2023)083}}.

\bibitem{Peet:1998wn}
A.~W. Peet, J.~Polchinski, {UV / IR relations in AdS dynamics}, Phys. Rev. D 59 (1999) 065011.
\newblock \href {http://arxiv.org/abs/hep-th/9809022} {\path{arXiv:hep-th/9809022}}, \href {https://doi.org/10.1103/PhysRevD.59.065011} {\path{doi:10.1103/PhysRevD.59.065011}}.

\bibitem{Papadimitriou:2004ap}
I.~Papadimitriou, K.~Skenderis, {AdS / CFT correspondence and geometry}, IRMA Lect. Math. Theor. Phys. 8 (2005) 73--101.
\newblock \href {http://arxiv.org/abs/hep-th/0404176} {\path{arXiv:hep-th/0404176}}, \href {https://doi.org/10.4171/013-1/4} {\path{doi:10.4171/013-1/4}}.

\bibitem{Bilson:2008ab}
S.~Bilson, {Extracting spacetimes using the AdS/CFT conjecture}, JHEP 08 (2008) 073.
\newblock \href {http://arxiv.org/abs/0807.3695} {\path{arXiv:0807.3695}}, \href {https://doi.org/10.1088/1126-6708/2008/08/073} {\path{doi:10.1088/1126-6708/2008/08/073}}.

\bibitem{Bilson:2010ff}
S.~Bilson, {Extracting spacetimes using the AdS/CFT conjecture: part II}, JHEP 02 (2011) 050.
\newblock \href {http://arxiv.org/abs/1012.1812} {\path{arXiv:1012.1812}}, \href {https://doi.org/10.1007/JHEP02(2011)050} {\path{doi:10.1007/JHEP02(2011)050}}.

\bibitem{Hubeny:2014qwa}
V.~E. Hubeny, {Covariant residual entropy}, JHEP 09 (2014) 156.
\newblock \href {http://arxiv.org/abs/1406.4611} {\path{arXiv:1406.4611}}, \href {https://doi.org/10.1007/JHEP09(2014)156} {\path{doi:10.1007/JHEP09(2014)156}}.

\bibitem{Headrick:2014eia}
M.~Headrick, R.~C. Myers, J.~Wien, {Holographic holes and differential entropy}, JHEP 10 (2014) 149.
\newblock \href {http://arxiv.org/abs/1408.4770} {\path{arXiv:1408.4770}}, \href {https://doi.org/10.1007/JHEP10(2014)149} {\path{doi:10.1007/JHEP10(2014)149}}.

\bibitem{Myers:2014jia}
R.~C. Myers, J.~Rao, S.~Sugishita, {Holographic holes in higher dimensions}, JHEP 06 (2014) 044.
\newblock \href {http://arxiv.org/abs/1403.3416} {\path{arXiv:1403.3416}}, \href {https://doi.org/10.1007/JHEP06(2014)044} {\path{doi:10.1007/JHEP06(2014)044}}.

\bibitem{Czech:2014wka}
B.~Czech, X.~Dong, J.~Sully, {Holographic reconstruction of general bulk surfaces}, JHEP 11 (2014) 015.
\newblock \href {http://arxiv.org/abs/1406.4889} {\path{arXiv:1406.4889}}, \href {https://doi.org/10.1007/JHEP11(2014)015} {\path{doi:10.1007/JHEP11(2014)015}}.

\bibitem{Balasubramanian:2018uus}
V.~Balasubramanian, C.~Rabideau, {The dual of non-extremal area: differential entropy in higher dimensions}, JHEP 09 (2020) 051.
\newblock \href {http://arxiv.org/abs/1812.06985} {\path{arXiv:1812.06985}}, \href {https://doi.org/10.1007/JHEP09(2020)051} {\path{doi:10.1007/JHEP09(2020)051}}.

\bibitem{Park:2022fqy}
C.~Park, C.-O. Hwang, K.~Cho, S.-J. Kim, {Dual geometry of entanglement entropy via deep learning}, Phys. Rev. D 106~(10) (2022) 106017.
\newblock \href {http://arxiv.org/abs/2205.04445} {\path{arXiv:2205.04445}}, \href {https://doi.org/10.1103/PhysRevD.106.106017} {\path{doi:10.1103/PhysRevD.106.106017}}.

\bibitem{Park:2023slm}
C.~Park, S.~Kim, J.~H. Lee, {Holography transformer} (11 2023).
\newblock \href {http://arxiv.org/abs/2311.01724} {\path{arXiv:2311.01724}}.

\bibitem{Ahn:2024jkk}
B.~Ahn, H.-S. Jeong, K.-Y. Kim, K.~Yun, {Holographic reconstruction of black hole spacetime: machine learning and entanglement entropy}, JHEP 01 (2025) 025.
\newblock \href {http://arxiv.org/abs/2406.07395} {\path{arXiv:2406.07395}}, \href {https://doi.org/10.1007/JHEP01(2025)025} {\path{doi:10.1007/JHEP01(2025)025}}.

\bibitem{Ahn:2024gjf}
B.~Ahn, H.-S. Jeong, K.-Y. Kim, K.~Yun, {Deep learning bulk spacetime from boundary optical conductivity}, JHEP 03 (2024) 141.
\newblock \href {http://arxiv.org/abs/2401.00939} {\path{arXiv:2401.00939}}, \href {https://doi.org/10.1007/JHEP03(2024)141} {\path{doi:10.1007/JHEP03(2024)141}}.

\bibitem{Jokela:2025ime}
N.~Jokela, T.~Liimatainen, M.~Sarkkinen, L.~Tzou, {Bulk metric reconstruction from entanglement data via minimal surface area variations} (4 2025).
\newblock \href {http://arxiv.org/abs/2504.07016} {\path{arXiv:2504.07016}}.

\bibitem{Wang:2018vbw}
P.~Wang, H.~Wu, H.~Yang, {Fixing three dimensional geometries from entanglement entropies of CFT$_{2}$}, Chin. Phys. C 49~(2) (2025) 025106.
\newblock \href {http://arxiv.org/abs/1809.01355} {\path{arXiv:1809.01355}}, \href {https://doi.org/10.1088/1674-1137/ad93b8} {\path{doi:10.1088/1674-1137/ad93b8}}.

\bibitem{Agon:2020mvu}
C.~A. Ag\'on, E.~C\'aceres, J.~F. Pedraza, {Bit threads, Einstein\textquoteright{}s equations and bulk locality}, JHEP 01 (2021) 193.
\newblock \href {http://arxiv.org/abs/2007.07907} {\path{arXiv:2007.07907}}, \href {https://doi.org/10.1007/JHEP01(2021)193} {\path{doi:10.1007/JHEP01(2021)193}}.

\bibitem{Alexakis2020}
S.~Alexakis, T.~Balehowsky, A.~Nachman, {Determining a Riemannian metric from minimal areas}, Advances in Mathematics 366 (2020) 107025.

\bibitem{Bao:2020abm}
N.~Bao, C.~Cao, S.~Fischetti, J.~Pollack, Y.~Zhong, {More of the bulk from extremal area variations}, Class. Quant. Grav. 38~(4) (2021) 047001.
\newblock \href {http://arxiv.org/abs/2009.07850} {\path{arXiv:2009.07850}}, \href {https://doi.org/10.1088/1361-6382/abcfd0} {\path{doi:10.1088/1361-6382/abcfd0}}.

\bibitem{Hammersley:2006cp}
J.~Hammersley, {Extracting the bulk metric from boundary information in asymptotically AdS spacetimes}, JHEP 12 (2006) 047.
\newblock \href {http://arxiv.org/abs/hep-th/0609202} {\path{arXiv:hep-th/0609202}}, \href {https://doi.org/10.1088/1126-6708/2006/12/047} {\path{doi:10.1088/1126-6708/2006/12/047}}.

\bibitem{Hammersley:2007ab}
J.~Hammersley, {Numerical metric extraction in AdS/CFT}, Gen. Rel. Grav. 40 (2008) 1619--1652.
\newblock \href {http://arxiv.org/abs/0705.0159} {\path{arXiv:0705.0159}}, \href {https://doi.org/10.1007/s10714-007-0564-6} {\path{doi:10.1007/s10714-007-0564-6}}.

\bibitem{Spillane:2013mca}
M.~Spillane, {Constructing space from entanglement entropy} (11 2013).
\newblock \href {http://arxiv.org/abs/1311.4516} {\path{arXiv:1311.4516}}.

\bibitem{Bao:2019bib}
N.~Bao, C.~Cao, S.~Fischetti, C.~Keeler, {Towards bulk metric reconstruction from extremal area variations}, Class. Quant. Grav. 36~(18) (2019) 185002.
\newblock \href {http://arxiv.org/abs/1904.04834} {\path{arXiv:1904.04834}}, \href {https://doi.org/10.1088/1361-6382/ab377f} {\path{doi:10.1088/1361-6382/ab377f}}.

\bibitem{Jokela:2020auu}
N.~Jokela, A.~P{\"o}nni, {Towards precision holography}, Phys. Rev. D 103~(2) (2021) 026010.
\newblock \href {http://arxiv.org/abs/2007.00010} {\path{arXiv:2007.00010}}, \href {https://doi.org/10.1103/PhysRevD.103.026010} {\path{doi:10.1103/PhysRevD.103.026010}}.

\bibitem{Cao:2020uvb}
C.~Cao, X.-L. Qi, B.~Swingle, E.~Tang, {Building bulk geometry from the tensor Radon transform}, JHEP 12 (2020) 033.
\newblock \href {http://arxiv.org/abs/2007.00004} {\path{arXiv:2007.00004}}, \href {https://doi.org/10.1007/JHEP12(2020)033} {\path{doi:10.1007/JHEP12(2020)033}}.

\bibitem{Jokela:2023rba}
N.~Jokela, K.~Rummukainen, A.~Salami, A.~P\"onni, T.~Rindlisbacher, {Progress in the lattice evaluation of entanglement entropy of three-dimensional Yang-Mills theories and holographic bulk reconstruction}, JHEP 12 (2023) 137.
\newblock \href {http://arxiv.org/abs/2304.08949} {\path{arXiv:2304.08949}}, \href {https://doi.org/10.1007/JHEP12(2023)137} {\path{doi:10.1007/JHEP12(2023)137}}.

\bibitem{Jafferis:2015del}
D.~L. Jafferis, A.~Lewkowycz, J.~Maldacena, S.~J. Suh, {Relative entropy equals bulk relative entropy}, JHEP 06 (2016) 004.
\newblock \href {http://arxiv.org/abs/1512.06431} {\path{arXiv:1512.06431}}, \href {https://doi.org/10.1007/JHEP06(2016)004} {\path{doi:10.1007/JHEP06(2016)004}}.

\bibitem{Dong:2016eik}
X.~Dong, D.~Harlow, A.~C. Wall, {Reconstruction of bulk operators within the entanglement wedge in gauge-gravity duality}, Phys. Rev. Lett. 117~(2) (2016) 021601.
\newblock \href {http://arxiv.org/abs/1601.05416} {\path{arXiv:1601.05416}}, \href {https://doi.org/10.1103/PhysRevLett.117.021601} {\path{doi:10.1103/PhysRevLett.117.021601}}.

\bibitem{Almheiri:2014lwa}
A.~Almheiri, X.~Dong, D.~Harlow, {Bulk locality and quantum error correction in AdS/CFT}, JHEP 04 (2015) 163.
\newblock \href {http://arxiv.org/abs/1411.7041} {\path{arXiv:1411.7041}}, \href {https://doi.org/10.1007/JHEP04(2015)163} {\path{doi:10.1007/JHEP04(2015)163}}.

\bibitem{Pastawski:2015qua}
F.~Pastawski, B.~Yoshida, D.~Harlow, J.~Preskill, {Holographic quantum error-correcting codes: Toy models for the bulk/boundary correspondence}, JHEP 06 (2015) 149.
\newblock \href {http://arxiv.org/abs/1503.06237} {\path{arXiv:1503.06237}}, \href {https://doi.org/10.1007/JHEP06(2015)149} {\path{doi:10.1007/JHEP06(2015)149}}.

\bibitem{Hayden:2016cfa}
P.~Hayden, S.~Nezami, X.-L. Qi, N.~Thomas, M.~Walter, Z.~Yang, {Holographic duality from random tensor networks}, JHEP 11 (2016) 009.
\newblock \href {http://arxiv.org/abs/1601.01694} {\path{arXiv:1601.01694}}, \href {https://doi.org/10.1007/JHEP11(2016)009} {\path{doi:10.1007/JHEP11(2016)009}}.

\bibitem{Roy:2018ehv}
S.~R. Roy, D.~Sarkar, {Bulk metric reconstruction from boundary entanglement}, Phys. Rev. D 98~(6) (2018) 066017.
\newblock \href {http://arxiv.org/abs/1801.07280} {\path{arXiv:1801.07280}}, \href {https://doi.org/10.1103/PhysRevD.98.066017} {\path{doi:10.1103/PhysRevD.98.066017}}.

\bibitem{Kabat:2018smf}
D.~Kabat, G.~Lifschytz, {Emergence of spacetime from the algebra of total modular Hamiltonians}, JHEP 05 (2019) 017.
\newblock \href {http://arxiv.org/abs/1812.02915} {\path{arXiv:1812.02915}}, \href {https://doi.org/10.1007/JHEP05(2019)017} {\path{doi:10.1007/JHEP05(2019)017}}.

\bibitem{Engelhardt:2016wgb}
N.~Engelhardt, G.~T. Horowitz, {Towards a reconstruction of general bulk metrics}, Class. Quant. Grav. 34~(1) (2017) 015004.
\newblock \href {http://arxiv.org/abs/1605.01070} {\path{arXiv:1605.01070}}, \href {https://doi.org/10.1088/1361-6382/34/1/015004} {\path{doi:10.1088/1361-6382/34/1/015004}}.

\bibitem{Engelhardt:2016crc}
N.~Engelhardt, G.~T. Horowitz, {Recovering the spacetime metric from a holographic dual}, Adv. Theor. Math. Phys. 21 (2017) 1635--1653.
\newblock \href {http://arxiv.org/abs/1612.00391} {\path{arXiv:1612.00391}}, \href {https://doi.org/10.4310/ATMP.2017.v21.n7.a2} {\path{doi:10.4310/ATMP.2017.v21.n7.a2}}.

\bibitem{Hernandez-Cuenca:2020ppu}
S.~Hern\'andez-Cuenca, G.~T. Horowitz, {Bulk reconstruction of metrics with a compact space asymptotically}, JHEP 08 (2020) 108.
\newblock \href {http://arxiv.org/abs/2003.08409} {\path{arXiv:2003.08409}}, \href {https://doi.org/10.1007/JHEP08(2020)108} {\path{doi:10.1007/JHEP08(2020)108}}.

\bibitem{Caron-Huot:2022lff}
S.~Caron-Huot, {Holographic cameras: an eye for the bulk}, JHEP 03 (2023) 047.
\newblock \href {http://arxiv.org/abs/2211.11791} {\path{arXiv:2211.11791}}, \href {https://doi.org/10.1007/JHEP03(2023)047} {\path{doi:10.1007/JHEP03(2023)047}}.

\bibitem{Hashimoto:2020mrx}
K.~Hashimoto, {Building bulk from Wilson loops}, PTEP 2021~(2) (2021) 023B04.
\newblock \href {http://arxiv.org/abs/2008.10883} {\path{arXiv:2008.10883}}, \href {https://doi.org/10.1093/ptep/ptaa183} {\path{doi:10.1093/ptep/ptaa183}}.

\bibitem{Hashimoto:2021umd}
K.~Hashimoto, R.~Watanabe, {Bulk reconstruction of metrics inside black holes by complexity}, JHEP 09 (2021) 165.
\newblock \href {http://arxiv.org/abs/2103.13186} {\path{arXiv:2103.13186}}, \href {https://doi.org/10.1007/JHEP09(2021)165} {\path{doi:10.1007/JHEP09(2021)165}}.

\bibitem{Czech:2014ppa}
B.~Czech, L.~Lamprou, {Holographic definition of points and distances}, Phys. Rev. D 90 (2014) 106005.
\newblock \href {http://arxiv.org/abs/1409.4473} {\path{arXiv:1409.4473}}, \href {https://doi.org/10.1103/PhysRevD.90.106005} {\path{doi:10.1103/PhysRevD.90.106005}}.

\bibitem{Czech:2014tva}
B.~Czech, P.~Hayden, N.~Lashkari, B.~Swingle, {The information theoretic interpretation of the length of a curve}, JHEP 06 (2015) 157.
\newblock \href {http://arxiv.org/abs/1410.1540} {\path{arXiv:1410.1540}}, \href {https://doi.org/10.1007/JHEP06(2015)157} {\path{doi:10.1007/JHEP06(2015)157}}.

\bibitem{Burda:2018rpb}
P.~Burda, R.~Gregory, A.~Jain, {Holographic reconstruction of bubble spacetimes}, Phys. Rev. D 99~(2) (2019) 026003.
\newblock \href {http://arxiv.org/abs/1804.05202} {\path{arXiv:1804.05202}}, \href {https://doi.org/10.1103/PhysRevD.99.026003} {\path{doi:10.1103/PhysRevD.99.026003}}.

\bibitem{Balasubramanian:2013lsa}
V.~Balasubramanian, B.~D. Chowdhury, B.~Czech, J.~de~Boer, M.~P. Heller, {Bulk curves from boundary data in holography}, Phys. Rev. D 89~(8) (2014) 086004.
\newblock \href {http://arxiv.org/abs/1310.4204} {\path{arXiv:1310.4204}}, \href {https://doi.org/10.1103/PhysRevD.89.086004} {\path{doi:10.1103/PhysRevD.89.086004}}.

\bibitem{Vidal:2014aal}
G.~Vidal, Y.~Chen, {Entanglement contour}, J. Stat. Mech. 2014~(10) (2014) P10011.
\newblock \href {http://arxiv.org/abs/1406.1471} {\path{arXiv:1406.1471}}, \href {https://doi.org/10.1088/1742-5468/2014/10/P10011} {\path{doi:10.1088/1742-5468/2014/10/P10011}}.

\bibitem{Bhattacharya:2012zu}
J.~Bhattacharya, S.~Cremonini, A.~Sinkovics, {On the IR completion of geometries with hyperscaling violation}, JHEP 02 (2013) 147.
\newblock \href {http://arxiv.org/abs/1208.1752} {\path{arXiv:1208.1752}}, \href {https://doi.org/10.1007/JHEP02(2013)147} {\path{doi:10.1007/JHEP02(2013)147}}.

\bibitem{Kundu:2012jn}
N.~Kundu, P.~Narayan, N.~Sircar, S.~P. Trivedi, {Entangled dilaton dyons}, JHEP 03 (2013) 155.
\newblock \href {http://arxiv.org/abs/1208.2008} {\path{arXiv:1208.2008}}, \href {https://doi.org/10.1007/JHEP03(2013)155} {\path{doi:10.1007/JHEP03(2013)155}}.

\bibitem{Bachas:2020yxv}
C.~Bachas, S.~Chapman, D.~Ge, G.~Policastro, {Energy reflection and transmission at 2D holographic interfaces}, Phys. Rev. Lett. 125~(23) (2020) 231602.
\newblock \href {http://arxiv.org/abs/2006.11333} {\path{arXiv:2006.11333}}, \href {https://doi.org/10.1103/PhysRevLett.125.231602} {\path{doi:10.1103/PhysRevLett.125.231602}}.

\bibitem{Simidzija:2020ukv}
P.~Simidzija, M.~Van~Raamsdonk, {Holo-ween}, JHEP 12 (2020) 028.
\newblock \href {http://arxiv.org/abs/2006.13943} {\path{arXiv:2006.13943}}, \href {https://doi.org/10.1007/JHEP12(2020)028} {\path{doi:10.1007/JHEP12(2020)028}}.

\bibitem{Liu:2024oxg}
Y.~Liu, H.-D. Lyu, C.-Y. Wang, {On AdS$_3$/ICFT$_2$ with a dynamical scalar field located on the brane} (3 2024).
\newblock \href {http://arxiv.org/abs/2403.20102} {\path{arXiv:2403.20102}}.

\bibitem{Liu:2025gle}
Y.~Liu, C.-Y. Wang, Y.-J. Zeng, {Energy transport in holographic non-conformal interfaces} (3 2025).
\newblock \href {http://arxiv.org/abs/2503.20399} {\path{arXiv:2503.20399}}.

\end{thebibliography}
\end{document}